\documentclass[preprint,3p,12pt]{elsarticle}
\usepackage{mathrsfs}
\usepackage{amsmath}
\usepackage{stmaryrd}
\usepackage{bbding}
\usepackage{dcolumn}
\usepackage{graphicx}
\usepackage{amsfonts}
\usepackage{amssymb}
\usepackage{psfrag}
\usepackage{wrapfig}
\usepackage{subfigure}
\usepackage{makeidx}
\usepackage{bm}
\usepackage{epsf}
\usepackage{epsfig}
\usepackage{setspace}
\usepackage{graphicx}
\usepackage{epstopdf}
\usepackage{psfrag}
\usepackage{subfigure}
\usepackage{color}

\begin{document}
\title{High-order Compact Gas-kinetic Schemes for Three-dimensional Flow Simulation on Tetrahedral Mesh}

\author[HKUST1]{Fengxiang Zhao}
\ead{fzhaoac@connect.ust.hk}

\author[HKUST2]{Xing Ji}
\ead{xjiad@connect.ust.hk}

\author[HKUST1]{Wei Shyy}
\ead{weishyy@ust.hk}

\author[HKUST1,HKUST2,HKUST3,HKUST4]{Kun Xu\corref{cor}}
\ead{makxu@ust.hk}

\address[HKUST1]{Department of Mechanical and Aerospace Engineering, Hong Kong University of Science and Technology, Clear Water Bay, Kowloon, HongKong}
\address[HKUST2]{Department of Mathematics, Hong Kong University of Science and Technology, Clear Water Bay, Kowloon, HongKong}
\address[HKUST3]{Center for Ocean Research in Hong Kong and Macau (CORE), Hong Kong University of Science and Technology, Clear Water Bay, Kowloon, Hong Kong}
\address[HKUST4]{Shenzhen Research Institute, Hong Kong University of Science and Technology, Shenzhen, China}
\cortext[cor]{Corresponding author}

\begin{abstract}

A general framework for the development of high-order compact schemes has been proposed recently.
The core steps of the schemes are composed of the following.
1). Based on a kinetic model equation, from a generalized initial distribution of flow variables construct a time-accurate evolution solution of gas distribution function at a cell interface ;
2). Introduce the WENO-type weighting functions into the time-derivative of the cell interface flux function in the multistage multi-derivative time stepping scheme to cope with the possible impingement of a shock wave on a cell interface within a time step;
3). Take moments of interface gas distribution function to obtain the time-accurate flow variables and the corresponding fluxes at the cell interface, and update the cell-averaged flow variables and their gradients inside each control volume;
4). Within the physical domain of dependence of the reconstructed cell, based on the cell-averaged flow variables and their gradients
develop compact initial data reconstruction to get initial flow distributions at the beginning of next time step.
A compact gas-kinetic scheme (GKS) up to sixth-order accuracy in space and fourth-order in time has been constructed on 2D unstructured mesh before.
In this paper, the compact GKS up to fourth-order accuracy on 3D tetrahedral mesh will be further constructed with the focus on the
WENO-type initial data reconstruction.
Nonlinear weights are designed to achieve high-order accuracy for the smooth Navier-Stokes solution and keep super robustness in 3D computation with strong shock interactions.
The fourth-order compact GKS can use a large time step with  CFL number $0.6$ in the simulations from subsonic to hypersonic flow.
A series of test cases are used to validate the scheme.
The high-order compact GKS is ready for 3D applications with complex geometry.

\end{abstract}

\begin{keyword}
Compact scheme, High-order GKS, WENO reconstruction, Tetrahedral mesh
\end{keyword}

\maketitle

\section{Introduction}
Over the last decades, the development of high-order schemes with the order $\geq 3$
has received great attention in computational fluid dynamics (CFD) research community.
Compared with second-order scheme, high-order schemes show overwhelming advantages in efficiency and accuracy
in scale-resolved simulations \cite{wang2013-review,wang2021-AIA,wang2022-WMLES},
such as large eddy simulation (LES) and direct numerical simulation (DNS).
In flow problems with complex geometry, unstructured meshes are often used due to the flexibility and automatical mesh generation.
High-order compact schemes \cite{reed,cockburn1,cockburn2,qiu1,qiu3,dumbser_pnpm,luo2,FR,CPR_wang} are preferred on unstructured mesh due to the low storage, high resolution, and high parallel efficiency.
The use of compact stencils is crucially important on unstructured mesh, especially in three-dimensional flow computation.
For the above compact schemes, besides the cell-averaged flow variables, additional variables or degrees of freedom (DOFs) are needed locally
to achieve high-order accuracy.

Different from the compact schemes with the Riemann solver as the underlying dynamic model and Runge-Kutta time stepping for the temporal evolution, the compact GKS is based on a multidimensional evolution solution, i.e., the time-accurate gas distribution function, to evaluate both flow variables and their corresponding fluxes on the cell interface.
As a result, the cell-averaged conservative flow variables and their gradients inside each control volume can be updated to the next time level and be used in the initial data reconstruction with compact stencils.
The high-order compact GKS has been developed in two-dimensional space with triangular mesh \cite{zhao2021direct}.
Besides the adoption of WENO-type methodology on the initial data reconstruction, in order to keep the robustness of the scheme and the use of a large time step, a nonlinear limiter through WENO formulation has been introduced into the high-order multistage multi-derivative time evolution process as well to cope with the possible discontinuous solution passing through a cell interface within a time step.
This kind of possible scene doesn't appear in the 1st-order evolution model of Riemann solver.
This framework makes the spatial and temporal discretization on an equal footing in the construction of high-order compact scheme.
Even with an accuracy up to 8th-order, the CFL number used in the compact GKS can take a value CFL$>0.8$
in two-dimensional computation on triangular mesh.
The numerical domain of  dependence matches with the physical domain of dependence in the compact scheme very well,
which clearly indicates the numerical modeling in the algorithm to be consistent with the space-time coupled fluid evolution with limited wave propagating speed.
In this paper, a compact GKS up to 4th-order accuracy on three-dimensional tetrahedral mesh will be constructed.
Different from the weak formulation in the update of flow variable at solution points in CPR \cite{CPR_wang,zhang2022-CPR,witherden2014-cpr},
the updating cell-averaged gradients through Gauss's theorem in GKS are based on the physical evolution solution with possible discontinuities at cell interfaces \cite{zhao2021direct}. Within the physical domain of dependence, the compact stencils provide the reconstructions in different orders.
In this paper, a simplified WENO method is adopted \cite{zhu2016-WENO,zhao_compact-tri}, with improved adaptivity for both smooth and discontinuous solutions. The nonlinear reconstruction is based on a combination of zeroth-order, first-order and higher-order polynomials, with the achievement of an optimal order of accuracy. With the implementation of simplified WENO reconstruction, the nonlinear compact GKS can present similar solution as that from the corresponding linear schemes in the well-resolved flow region, and capture discontinuous solution automatically.

This paper is organized as follows. The gas-kinetic evolution model of GKS will be introduced in Section 2. Section 3 is about the direct modeling of flow evolution at a cell interface for the update of cell-averaged conservative flow variables and their gradients.
Section 4 and Section 5 will present the linear and nonlinear compact reconstructions for the determination of piecewise high-order polynomials inside each control volume. In Section 6, the compact GKS will be tested in a wide range of cases in three-dimensional space on tetrahedral mesh. The last section is the conclusion.

\section{Time-accurate gas-kinetic evolution model}
For the construction of high-order compact GKS, the use of high-order gas evolution model beyond the 1st-order Riemann solution is necessary.
In this section, we will briefly  present the time-accurate evolution solution of a gas distribution at a cell interface.

The gas-kinetic evolution is based on the kinetic BGK equation \cite{BGK-1},
\begin{equation}\label{bgk}
f_t+\textbf{u}\cdot\nabla f=\frac{g-f}{\tau},
\end{equation}
where $\textbf{u}=(u,v,w)$ is the particle velocity, $f$ is the gas distribution function, $g$ is the corresponding equilibrium state that $f$ approaches, and $\tau$ is particle collision time.
The equilibrium state $g$ is a Maxwellian distribution,
\begin{equation*}
g=\rho(\frac{\lambda}{\pi})^{\frac{K+3}{2}}e^{-\lambda((\mathbf{u}-\mathbf{U})^2+\xi^2)},
\end{equation*}
where $\lambda =1/2RT $, and $R$ and $T$ are the gas constant and temperature, respectively.
$K$ is the number of internal DOFs, i.e. $K=(5-3\gamma)/(\gamma-1)$ for three-dimensional flow,
and $\gamma$ is the specific heat ratio. $\xi$ is the internal variable with $\xi^2=\xi^2_1+\xi^2_2+...+\xi^2_K$.
At a relatively low temperature without exciting the vibrational mode, a diatomic molecule in a three-dimensional flow has two rotational DOFs in $\xi$, such as $K=2$.
$\mathbf{U}=(U,V,W)$ is the macroscopic flow velocity which is the same velocity in the Navier-Stokes (NS) equations.
Due to the conservation of mass, momentum and energy during particle collisions, $f$ and $g$ satisfy the compatibility condition,
\begin{equation}\label{compatibility}
\int \frac{g-f}{\tau}\pmb{\psi} \mathrm{d}\Xi=0,
\end{equation}
at any point in space and time, where $\pmb{\psi}=(\psi_1,\psi_2,\psi_3,\psi_4,\psi_5)^T=(1,u,v,w,\displaystyle \frac{1}{2}(\textbf{u}^2+\xi^2))^T$, $\text{d}\Xi=\text{d}u\text{d}v\text{d}w\text{d}\xi_1...\text{d}\xi_{K}$.

The macroscopic conservative flow variables $\mathbf{W}=(\rho,\rho U,\rho V,\rho W,\rho E)$ can be evaluated from the gas distribution function,
\begin{equation}\label{g-to-convar}
\textbf{W}=\int f \pmb{\psi} \mathrm{d}\Xi.
\end{equation}
The corresponding fluxes for mass, momentum and energy in $i$-th direction is given by
\begin{equation}\label{g-to-flux}
{\textbf{F}_i} =\int u_i f \pmb{\psi} \mathrm{d}\Xi,
\end{equation}
with $(u_1,u_2,u_3)=\mathbf{u}$ in the 3-D case.
According to the Chapman$-$Enskog expansion \cite{xu1,xu2,xu2021unified}, the zeroth-order truncation $f=g$ corresponds to the invicid Euler equations, and the first-order truncation $f=g-\tau(\mathbf{u}\cdot \nabla_{\mathbf{u}}g+g_t)$ gives the NS equations.

In GKS, the evolution solution ${\bf W}(t)$ and ${\bf F}(t)$ at a cell interface are determined by the time-accurate gas distribution function $f$.
Based on the integral solution of BGK equation and the modeling for the initial state and equilibrium state distribution in local space and time \cite{xu2}, the time-accurate distribution function at a cell interface becomes
\begin{equation}\label{integral1}
\begin{split}
f(\textbf{x}_0,t,\textbf{u},\xi)=&\frac{1}{\tau}\int_0^t g(\textbf{x}',t',\textbf{u},\xi)e^{-(t-t')/\tau}\mathrm{d}t' \\
&+e^{-t/\tau}f_0(\textbf{x}_0-\textbf{u}(t-t_0),\textbf{u},\xi),
\end{split}
\end{equation}
where $\textbf{x}_0$ is the numerical quadrature point at the cell interface, and $\textbf{x}_0=\textbf{x}^{'}+\textbf{u}(t-t^{'})$ is the particle trajectory.
Here $f_0$ is the initial state of gas distribution function $f$ at $t=0$.
In order to explicitly obtain the solution $f$ of Eq.(\ref{integral1}), both $f_0$ and $g$ in Eq.(\ref{integral1}) need to be modeled.
The second-order accurate solution for $f$ is \cite{xu2}
\begin{align}\label{2nd-f}
\begin{split}
f(\textbf{x}_0,t,\textbf{u},\xi)=&(1-e^{-t/\tau})g_0+((t+\tau)e^{-t/\tau}-\tau)(\overline{a}_1u+\overline{a}_2v+\overline{a}_3w)g_0 \\
+&(t-\tau+\tau e^{-t/\tau}){\bar{A}} g_0 \\
+&e^{-t/\tau}g_l[1-(\tau+t)(a_{1l}u+a_{2l}v+a_{3l}w)-\tau A_l]H(u) \\
+&e^{-t/\tau}g_r[1-(\tau+t)(a_{1r}u+a_{2r}v+a_{3r}w)-\tau A_r](1-H(u)),
\end{split}
\end{align}
where the terms related to $g_0$ are from the integral of the equilibrium state and the terms related to $g_l$ and $g_r$ are from the initial term $f_0$ in the Eq.(\ref{integral1}). All the coefficients in Eq.(\ref{2nd-f}), such as $a_{kl}$ and $a_{kr}~(k=1,2,3)$, can be determined from the initially reconstructed macroscopic flow variables at the left and right sides of the cell interface.
The above time evolution solution is distinguishable from the generalized Riemann problem (GRP) \cite{grp-2006} in the following aspects.
(1). The above distribution function takes a physical process from the flux vector splitting transport to the NS solution at the cell interface $\textbf{x}_0$; (2). The flow evolution has multidimensional mechanism with the participation of $\partial x_i$ terms in the solution; (3). In smooth flow region, the evolution solution has the Lax-Wendroff type central difference property with involving and modeling of equilibrium state in the time-accurate solution; (4). The solution is obtained explicitly without iterations.

\section{Solution updates}

The discrete conservation law in a control volume $\Omega_j$ is,
\begin{equation}\label{conservation}
\int_{\Omega_j} {\bf W} ({\bf x}, t^{n+1} ) \mathrm{d}V = \int_{\Omega_j} {\bf W} ({\bf x}, t^{n} ) \mathrm{d}V -
\int_{t^n}^{t^{n+1}} \int_{\partial \Omega_j} {\bf F}(t) \cdot {\bf n} \mathrm{d}S \mathrm{d}t,
\end{equation}
where $\mathbf{W}(\mathbf{x},t)$ is the conservative flow variable in a control volume $\Omega_j$, and $\mathbf{F}(t)$ is the corresponding flux across the cell interface $\partial \Omega_j$.
The above integral conservation law is valid in any flow regimes from the rarefied to the continuum one once the dynamics of ${\bf F}(t)$ can be  properly modeled.
The accuracy of the updated solution depends critically on the time-dependent interface flux function ${\bf F}(t)$,
which depends on the initial condition ${\bf W}_j(t^n)$ and the evolution model for the cell interface flux ${\bf F}(t)$.

As a compact scheme, with the time-accurate interface flow variable ${\bf W}(t)$, the cell-averaged gradient of the flow variable can be updated as well by the Gauss's law,
\begin{equation}\label{slope}
 {\Omega_j} \nabla {\bf W}_j(t^{n+1}) =\int _{\Omega_j} \nabla {\bf W} ({\bf x}, t^{n+1} ) \mathrm{d}V =
 \int_{\partial \Omega_j} {\bf W}(t^{n+1}) {\bf n} \mathrm{d} S.
\end{equation}
With the consideration of possible discontinuous flow distribution around the cell interface,
the above flow variable $\mathbf{W}(t^{n+1})$ is the value at the inner side on the cell interface of the control volume.
In other words, ${\bf W}(\mathbf{x}_0, t^{n+1})$ may have multiple values, such as ${\bf W}^l(\mathbf{x}_0)$ and ${\bf W}^r(\mathbf{x}_0)$ at both sides of a cell interface. The outstanding example is that a stationary shock is exactly standing on the cell interface.
The evolution solution of ${\bf F}(t)$ and ${\bf W}^{l,r}(\mathbf{x}_0)$ will be presented next.

\subsection{Update of cell-averaged flow variable}

The conservation law in Eq.(\ref{conservation}) can be written as
\begin{equation}\label{semifvs}
\textbf{W}^{n+1}_{j}=\textbf{W}^{n}_{j} +\int_{t^n}^{t^{n+1}} {\cal L}_j (t) \mathrm{d} t,
\end{equation}
where $\textbf{W}_{j}$ is the cell-averaged flow variable defined as
\begin{align}\label{cell-average}
\textbf{W}_{j}=\frac{1}{\big| \Omega_j \big|} \int_{\Omega_j} \textbf{W}(\mathbf{x}) \text{d}V.
\end{align}
The surface integral in ${\cal L}_j (t)$ is discretized by a q-point Gaussian quadrature rule,
\begin{align}\label{semifvs-rhs}
{\cal L}_j (t)=-\frac{1}{|\Omega_j|}\int_{\partial \Omega_j} \textbf{F}\cdot \textbf{n} \mathrm{d} S=-\frac{1}{|\Omega_j|} \sum_{l=1}^{l_0}\big( \sum _{k=1}^q \omega_k \textbf{F}(\mathbf{x}_k)\cdot \textbf{n}_l \big) \big|\Gamma_{l} \big|,
\end{align}
where $\big|\Gamma_{l} \big|$ is the face area of the cell, $l_0$ is the number of cell faces, $\textbf{n}_l$ is the unit outer normal vector, and $q$ and $\omega_k$ are the number of quadrature points and weight of the Gaussian quadrature rule.

With the consideration of possible discontinuous flux function $\textbf{F}$ across the cell interface, such as
a moving shock with speed $v_s$ passing through the cell interface $v_s=[\textbf{F}^n]/[\textbf{W}^n]$, the flux function $\textbf{F}^n(t)$ on a cell interface may be a discontinuous function of time, where $\textbf{F}^n$ and $\textbf{W}^n$ are the normal components of $\textbf{F}$ and $\textbf{W}$, respectively. In order to capture such a dynamic evolution without introducing oscillation, same as nonlinear reconstruction polynomial of flow variables in space, the nonlinearly limited flux function in time has be developed as well.
Using a fourth-order time-accurate flux function in which terms of quadratic and above are limited, Eq.(\ref{semifvs}) is discretized as
\begin{equation}\label{S2O4}
\begin{split}
\mathbf{W}_j^{n+1/2} =&\mathbf{W}_j^n +\frac{\Delta t}{2}\mathcal{L}_{j}(\mathbf{W}^n) + \frac{\Delta t^2}{8}\mathcal{L}_{j,t}(\mathbf{W}^n), \\
\mathbf{W}_j^{n+1} =&\mathbf{W}^n +\Delta t\mathcal{L}_{j}(\mathbf{W}^n) + \frac{\Delta t^2}{2}\mathcal{L}_{j,t}(\mathbf{W}^n) \\
           &-\frac{\Delta t^2}{3}\widetilde{\mathcal{L}}_{j,t}(\mathbf{W}^n)+\frac{\Delta t^2}{3}\widetilde{\mathcal{L}}_{j,t}(\mathbf{W}^{n+1/2}),
\end{split}
\end{equation}
where at a middle step $t=t^{n+1/2}$ a nonlinearly limited flux function is introduced \cite{zhao2021direct},
$\widetilde{\mathcal{L}}_{j}$ is given as
\begin{equation}\label{semifvs-RHS-nonlinear}
\widetilde{\mathcal{L}}_{j}(\textbf{W}) = -\frac{1}{\big|\Omega_j\big|} \sum_{l=1}^{l_0} \omega^t_l \big( \sum _{k=1}^q \omega_k \textbf{F}(\mathbf{x}_k)\cdot \textbf{n}_l \big) \big|\Gamma_{l} \big|,
\end{equation}
where $\omega^t_l$ is a nonlinear weight for the $l$th interface of the cell and $\omega^t_l \in [0,1]$.
The nonlinear weight $\omega^t$ is defined as
\begin{equation}\label{S2O4-w}
\begin{split}
\widetilde{\alpha}^k_1&=1+ \big( \frac{\tau_Z}{IS^k_{s}+\epsilon}  \big)^{\alpha}, ~~\widetilde{\alpha}^k_2=1+ \big( \frac{\tau_Z}{IS^k_{d}+\epsilon}  \big)^{\alpha}, ~~k=L,R, \\
\alpha^k_2&=2\frac{\widetilde{\alpha}^k_2}{\widetilde{\alpha}^k_1+\widetilde{\alpha}^k_2}, \\
\omega^t&=\mathrm{min}\{\alpha^L_2,\alpha^R_2\},
\end{split}
\end{equation}
where $\alpha$ is a positive integer, and $\tau_Z$ is the local higher-order reference value to indicate smoothness of the large stencil in the spatial reconstruction.
$IS^{L,R}_{s}$ and $IS^{L,R}_{d}$ are smooth indicators corresponding a more smooth and a less smooth candidate polynomials in the nonlinear compact spatial reconstruction in the cells on both sides of the cell interface.
The determination of these parameters are given in Section 4.
The above fourth-order scheme in time can be reduced to a second-order one-step method for discretizing Eq.(\ref{semifvs}),
\begin{equation}\label{s1o2}
\mathbf{W}_j^{n+1} =\mathbf{W}_j^n +\Delta t \mathcal{L}_{j}(\mathbf{W}^n) + \frac{\Delta t^2}{2}\mathcal{L}_{j,t}(\mathbf{W}^n).
\end{equation}

\subsection{Update of cell-averaged gradient}

The RHS of Eq.(\ref{slope}) can be discretized by the same q-point Gaussian quadrature rule as that in Eq.(\ref{semifvs-rhs}).
The update of cell-averaged gradient becomes
\begin{align}\label{slope-1}
\nabla \textbf{W}_{j}& =\frac{1}{\big| \Omega_j \big|}\sum_{l=1}^{l_0} \big(|\Gamma_l| \mathbf{n}_{l} \sum _{k=1}^q \omega_k\textbf{W}^{n+1}(\textbf{x}_k) \big),
\end{align}
where $\textbf{W}^{n+1}(\textbf{x}_k)$ is the value at the inner surface of $\Omega_j$, which may be different from the value at the other side of the surface. In order to obtain a high-order accurate flow variable at the quadrature point, \
the macroscopic flow variable is evolved by two stages
\begin{equation}\label{S2O3}
\begin{split}
\textbf{W}^{n+1/2}(\textbf{x})=\textbf{W}^n(\textbf{x})+\frac{1}{2}\Delta t \textbf{W}_{t}^n(\textbf{x}), \\
\textbf{W}^{n+1}(\textbf{x})  =\textbf{W}^n(\textbf{x})+\Delta t \textbf{W}_{t}^{n+1/2}(\textbf{x}).
\end{split}
\end{equation}
To provide the flow variable at both sides of a cell interface \cite{zhao2021direct},
the update model for $\textbf{W}(\mathbf{x},t)$ is given by
\begin{align}\label{Gauss-theorem}
\begin{split}
{\bf W}^l(x,t)&= (1-e^{-\Delta t/\tau_0}){\bf W}^e(x,t) +e^{-\Delta t/\tau_0} {\bf W}_0^l(x,t), \\
{\bf W}^r(x,t)&= (1-e^{-\Delta t/\tau_0}){\bf W}^e(x,t) +e^{-\Delta t/\tau_0} {\bf W}_0^r(x,t).
\end{split}
\end{align}
The evolution solution ${\bf W}^e$ is given by the moments of the time-accurate distribution function in Eq.(\ref{2nd-f}),
and ${\bf W}_0^l$ and ${\bf W}_0^r$ are obtained from Eq.(\ref{2nd-f}) as well with the assumptions smooth initial condition on both
sides of the cell interface separately.
The weighting function $e^{-\Delta t/\tau_0}$ is constructed from a physical relaxation model, while
in the smooth flow region under $\Delta t \gg \tau_0$, a single flow variable at the cell interface is recovered.
The relaxation time $\tau_0$ is defined as
\begin{align*}
\tau_0=\varepsilon_{diss} \big|\frac{p_l-p_r}{p_l+p_r}\big|\Delta t,
\end{align*}
where $p_l$ and $p_r$ are the pressures at both sides of the cell interface,
and $\varepsilon_{diss}$ is a constant coefficient with a uniform value $\varepsilon_{diss}=5$ in all test cases in this paper.
The same modeling of relaxation time in case of numerical shock wave is firstly proposed in GKS \cite{xu2}.

\section{Stencils and linear compact reconstruction}

The compact linear reconstruction from second to fourth order of accuracy is given in this section.
Although the second-order scheme in one- and two-dimensional cases is well developed, the construction of a second-order scheme on tetrahedral mesh is not trivial due to its characteristics of the geometry.
The first-level neighbors of a tetrahedron may not fill up the space around it, as shown in Fig. \ref{1-stencil}, and the domain of influence to the central tetrahedron cannot be fully covered by the direct neighboring mesh.
As a result, on tetrahedral mesh with the stability requirement a second-order finite volume scheme should use a relative large stencils which include the first-level and second-level neighboring cells \cite{stability_unstructure}.
In order to present a complete picture for the reconstruction, a second-order compact reconstruction is also presented in this section.

\subsection{Compact reconstruction}

\begin{figure}[!htb]
\centering
\includegraphics[width=0.35\textwidth]{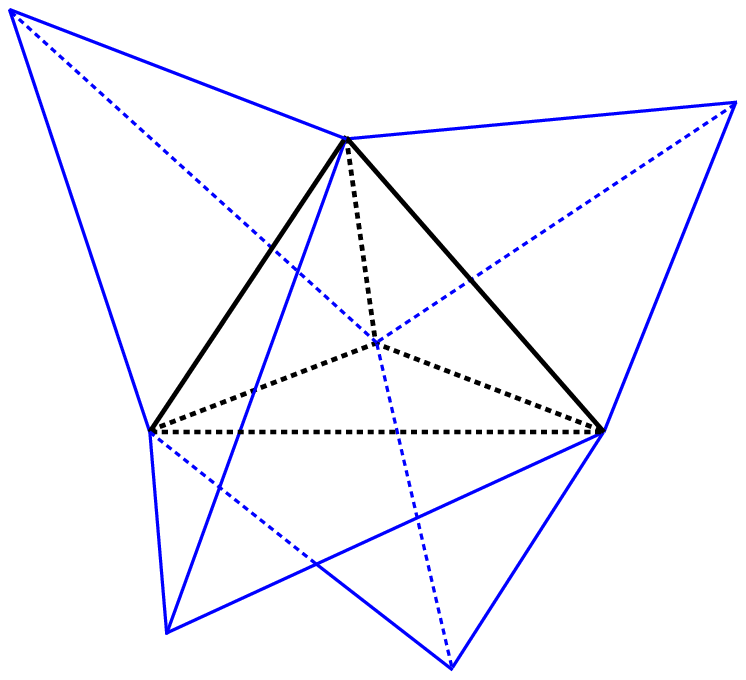}
\includegraphics[width=0.35\textwidth]{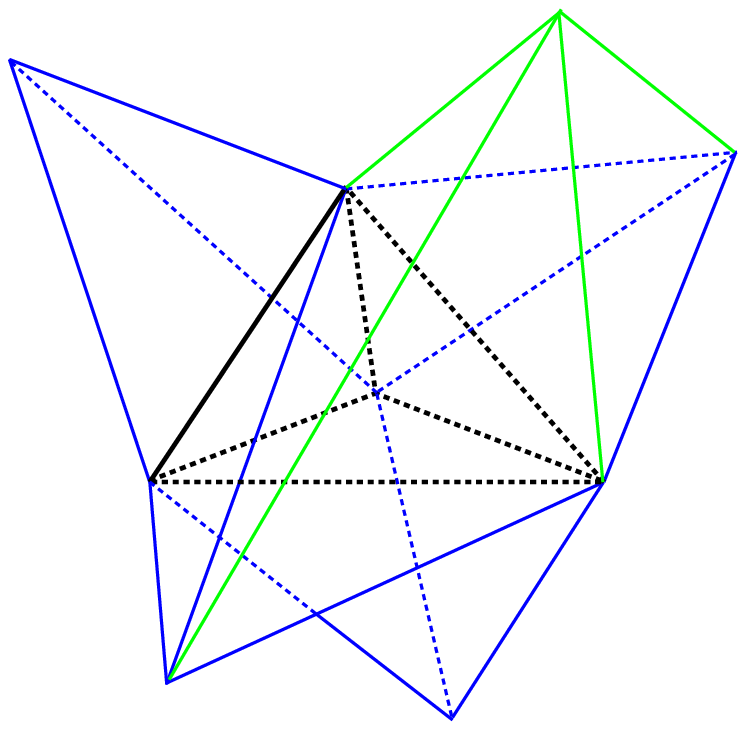}
\caption{\label{1-stencil} A schematic of reconstruction stencils of second-order to fourth-order compact GKS. The cell with black edges is the reconstructed cell. The cells with blue edges are the first-level neighbor cells of the reconstructed cell, and the cells with green edges are the second-level neighbors of the reconstructed cell. For the sake of simplicity and clarity of the illustration, only two cells of the second-level neighboring cells are shown. In general, a first-level cell is connected with three second-level neighboring cells with common nodes to the reconstructed cell. }
\end{figure}

A schematic of reconstruction stencils of second- to fourth-order compact GKS are shown in Fig. \ref{1-stencil}.
The cell with black edges is the reconstructed cell. The cells with blue edges are the first-level neighboring cells of the reconstructed cell, and the cells with green edges are the second-level neighbors of the reconstructed cell. For the sake of simplicity and clarity of illustration, only two cells among the second-level neighboring cells are shown. In general, a first-level neighboring cell is connected to three second-level neighboring cells.

A reasonable stencil consistent with the physical domain of dependence should consist of the reconstructed cell and all neighboring cells sharing common nodes with it.
Considering the simplicity of the algorithm and a finite order of accuracy in reconstruction,
many subset stencils with different sizes can be defined from the largest complete and compact stencil.
In the current compact GKS, inside each control volume, one cell-averaged flow variable and $3$ cell-averaged derivatives are available.
The stencils and adopted data for the compact reconstruction are determined from the following consideration.

\begin{enumerate}[(1)]
\item The compact stencils are the subsets of the largest compact set consisting of the reconstructed cell and its neighboring cells with common nodes.
\item For the $r$th-order compact reconstruction, the number of adopted data is about $N_p=2N_{DOF}$, where $N_{DOF}$ is the number of DOFs of the reconstructed $r$th-order polynomial.
\item The data for reconstruction is selected in the sequence from the reconstructed cell to the first-level neighbors and then to the second-level neighbors.
\item For the same level neighboring cells, the cell-averaged variable has a higher priority than the cell-averaged gradient.
\end{enumerate}

For the reconstructed cell $0$, the four first-level neighboring cells are denoted as $1$ to $4$.
In general, each first-level cell is connected to other three second-level neighboring cells through the surfaces,
and the numbers of the second-level cells for each first-level cell j $(j=1,2,3,4)$ are defined as $5+3(j-1), 6+3(j-1)$ and $7+3(j-1)$.
For the third- and fourth-order compact reconstruction, the stencils and adopted data can be obtained from
\begin{align*}
\begin{split}
S^2&=\{Q_0,Q_{j_1},\nabla Q_{0},\nabla Q_{j_1}\}, ~j_1=1,2,3,4, \\
S^3&=\{Q_0,Q_{j_1},Q_{j_2},\nabla Q_{0},\nabla Q_{j_1},\nabla_\mathbf{l} Q_{j_2}\}, ~j_1=1,2,3,4, ~j_2=5,6,\cdots,16,
\end{split}
\end{align*}
where $\nabla Q_{0}=(Q_{0,x},Q_{0,y},Q_{0,z})$ and $\nabla_\mathbf{l} Q_{j_2}$ is the directional derivative defined by
\begin{align*}
\nabla_\mathbf{l} Q_{j_2}\equiv \frac{\partial Q_{j_2}}{\partial \mathbf{l}}=\nabla Q_{j_2} \cdot \mathbf{n_l},
\end{align*}
where $\mathbf{n_l}$ is the unit vector along the direction $\bf{l}$.
Suppose second-level neighboring cell $\Omega_{j_2}$ is connected to first-level neighboring cell $\Omega_{j_1}$, and $\Omega_{j_1}$ and $\Omega_{0}$ is connected by face $\Gamma_{0-j_1}$.
For $\nabla_\mathbf{l} Q_{j_2}$, $\mathbf{n_l}$ is taken as the unit outer normal vector of the face $\Gamma_{0-j_1}$.

At the boundary of computational domain,
a first-level ghost cells are constructed based on the boundary condition.
But the next level neighboring cells of the first-level ghost cell are not constructed.
Therefore, the reconstruction scheme based on the stencils on the right figure of Fig. \ref{1-stencil} cannot be directly applied due to the absence of cells.
In this paper, a third-order compact reconstruction will be developed for the boundary cells and the scheme is stable in numerical examples.

Besides the high-order reconstruction,
to get back to the second-order nonlinear reconstruction is critical for the flow simulation at discontinuities.
Due to the use of the cell-averaged flow variables and their gradients,
the second-order reconstruction will be different from the traditional methods with limiters.
Biased stencils will be used instead of a central stencil to get second-order compact reconstruction.
Total four biased compact stencils for second-order reconstruction can be obtained
\begin{align*}
\begin{split}
S^1&=S^1_1\cup S^1_2\cup S^1_3 \cup S^1_4, S^1_k=\{Q_0,Q_{j},\nabla Q_{j},\}, ~j=1,2,3,4. \\
\end{split}
\end{align*}
Since the DOFs (zeroth-order and first-order derivatives) of a linear polynomial can be determined by the cell average and its cell-averaged gradient, the above biased stencils can fully determine linear polynomials.
Then, the linear combination of the four linear polynomials can give a second-order compact reconstruction.

\subsection{Linear system of compact reconstruction}
The polynomial function used in the compact reconstruction is written as
\begin{align*}
p^r(\mathbf{x})=Q_0+\sum_{k=1}^{N_{DOF}-1} a_k \varphi_k(\mathbf{x}),
\end{align*}
where $a_k$ is the DOF of the reconstruction polynomial. $\varphi_k$ is the zero-mean basis defined by
\begin{align*}
\begin{split}
&\varphi_k(\mathbf{x})=\frac{1}{l!m!n!}\delta x^l \delta y^m \delta z^n - \overline{\frac{1}{l!m!n!} \delta x^l \delta y^m \delta z^n}^{(0)}, \\
&\delta x=\frac{1}{h_x}(x-x_0), \delta x=\frac{1}{h_y}(y-y_0), \delta z=\frac{1}{h_z}(z-z_0),
\end{split}
\end{align*}
where $0\leq l,m,n\leq r$ and $~\mathrm{max}\{l+m+n\}=r$. $\varphi_k$ can make $p^r$ automatically satisfy the conservation condition in the reconstruction.
$h_x,~h_y$ and $h_z$ are the characteristic scales of $\Omega_0$ along the three directions of axes, which take the values
$h_x=h_y=h_z=h=|\Omega_j|^{1/3}$ for isotropic mesh.
To divide by $h$ in the expansion is to make the condition number of the matrix in the linear system of $a_k$ small.
$p^r(\mathbf{x})$ is constrained by the following conditions
\begin{align}\label{CLS-p1}
\begin{split}
\frac{1}{|\Omega_{j}|}\int_{\Omega_{j}} \varphi_k(\mathbf{x}) \mathrm{d}\mathbf{x} a_k&=Q_{j}, \\
\frac{h}{|\Omega_{j}|}\int_{\Omega_{j}} \varphi_{k,l}(\mathbf{x}) \mathrm{d}\mathbf{x} a_k&=hQ_{{j},l}, ~\Omega_{j}\in S^r,~l=x,y,z.
\end{split}
\end{align}
Based on the above constraints for $p^r(\mathbf{x})$, the linear system for $a_k$ is obtained by the least square (LS) method or constrained least square (CLS) method.
In the CLS method, some constraints, such as the one for $Q_{j}$ ($j=1,2,3,4$), are strictly satisfied, and others are satisfied in the sense of least square.

For $p^1$ and $p^2$, the CLS method is adopted, where the cell-averaged values $Q_{j}~(j=1,2,3,4)$ are strictly satisfied.
While LS method is adopted for $p^3$, which makes the fourth-order compact GKS have better stability on irregular meshes.
If the CLS method is adopted for $p^3$ and the cell-averaged values of the first-level neighbors are strictly satisfied, the linear system of $a_k$ is more sensitive  on the constraints from the first-level neighboring cells, due to the use of a smaller
effective numerical domain of dependence.
The CLS problem in the determination of $p^1$ and $p^2$ can be solved by the Lagrangian factor method and the linear system for $a_k$ can be obtained.
A general form of the linear system for $a_k$ has been given in \cite{zhao_compact-tri}.

The linear polynomial of the second-order reconstruction can be the linear combination of four $p^1_k(\mathbf{x})$.
\begin{equation}\label{p1-linear}
p^1(\mathbf{x})=\sum_{k=1}^4 \frac{1}{4}p^1_k(\mathbf{x}).
\end{equation}

\section{Nonlinear compact reconstruction}
To deal with discontinuities, nonlinear compact reconstruction with WENO method is adopted. WENO reconstruction can adaptively achieve high-order accuracy in smooth region and essentially non-oscillatory property in discontinuity region.
The simplified WENO method is developed for simple implementation and good robustness on unstructured mesh \cite{zhu2016-WENO,zhao_compact-tri}.
The basis of the adaptivity in the simplified WENO is the nonlinear combination of a high-order polynomial and several lower-order polynomials.
The four compact reconstructions $p^1_k~(k=1,2,3,4)$ on the biased stencils can be used as the lower-order candidate polynomials in the WENO method.
\subsection{Simplified WENO reconstruction}
The extension of the one-dimensional WENO reconstruction to unstructured mesh is difficult \cite{zhao2018}, especially for high-order (order $\geq4$) one. The simplified WENO reconstruction has been developed and implemented on triangular mesh \cite{zhao_compact-tri}.
The simplified WENO reconstruction is given as
\begin{equation}\label{weno}
\begin{split}
R(\mathbf{x})=\sum_{k=1}^{n}w_k p_k(\mathbf{x}) + w_0 \big( \frac{1+C}{C}p_0(\mathbf{x}) -\sum_{k=1}^{n} \frac{C_k}{C}p_k(\mathbf{x}) \big).
\end{split}
\end{equation}
The nonlinear weight $w_k$ is
\begin{equation}\label{weno-w}
\begin{split}
&w_{k}=\frac{ \widetilde{w}_{k} }{ \sum_{j=0}^{n} \widetilde{w}_{j}  }, \\
&\widetilde{w}_{k}=d_k \big( 1+ \big( \frac{\tau_Z}{IS_k+\epsilon}  \big)^{\alpha} \big),
\end{split}
\end{equation}
and the linear weight $d_{k}$ is
\begin{equation}\label{linear-dk}
d_{0}=\frac{C}{1+C}, ~~ d_k=\frac{C_k}{1+C},~~ k=1,\cdots,n.
\end{equation}
where $\epsilon$ is a small positive number and $\epsilon=1\times 10^{-15}$ is taken for all numerical tests in this paper, $n$ is the number of the sub-stencils.
A large number $\alpha$ improves the robustness of the scheme through better identifying the less smooth polynomials from all candidate polynomials. $\alpha=3$ is taken in the current compact reconstruction.
$\tau_Z$ is the local high-order reference value to indicate smoothness of the large stencil and it is given by $IS_k$, where $IS_k$ is obtained by the conventional definition in \cite{jiang-WENO,hu1999-WENO-tri}.
The parameters $C$ and $C_k$ are required to satisfy $\sum_{k=1}^{n} C_k=1, ~~ C>0$.
$C=n$ and $C_k=1/n$ are taken in this paper. The compact scheme based on the simplified WENO reconstruction is insensitive to the values of $C$ and $C_k$.

Based on the concept for reconstruction without crossing discontinuity of the solution, the biased lower-order polynomials $p^k$ in the WENO reconstruction take four biased linear polynomials $p^1_k$ and one zeroth-order polynomial $p^0$.
The smooth indicator of $p^0$ is obtained based on the local smoothest slope $W_{j,x},~W_{j,y}$ and $W_{j,z}$ ($j=0,1,2,3,4$) by the definition in \cite{hu1999-WENO-tri}. Such a construction guarantees that the smoothness indicator of $p^0$ corresponds to an auxiliary smoother linear reconstruction than linear reconstruction $p^1_k$.

The calculation of $IS$ based on the definition in \cite{hu1999-WENO-tri} will be complicated for the high-order polynomial (order $\geq 3$) on the tetrahedral mesh. In this paper, numerical quadrature is adopted to calculate $IS$. For a third-order polynomial $p^3$, its $IS$ can be
\begin{equation}\label{IS}
\begin{split}
IS &= \sum_{|\alpha|=1}^{3} h^{2|\alpha|-3} \int_{\Omega_{j}} \big( \frac{\partial^{|\alpha|} p^3(\mathbf{x})}{\partial x^{\alpha_1}\partial x^{\alpha_2}\partial x^{\alpha_3}} \big)^2dV \\
   &\approx \sum_{|\alpha|=1}^{2} h^{2|\alpha|} \big( \frac{\partial^{|\alpha|} p^3(\mathbf{x}_0)}{\partial x^{\alpha_1}\partial x^{\alpha_2}\partial x^{\alpha_3}} \big)^2,
\end{split}
\end{equation}
where $\mathbf{x}_0$ is the centroid of $\Omega_{j}$. $|\alpha|$ is a multi-index, for example, when $|\alpha|=1$, there are cases $(\alpha_1,\alpha_2,\alpha_3)=(1,0,0),~(0,1,0)$ and $(0,0,1)$. Apparently, two simplified calculations are made in Eq.(\ref{IS}). Firstly, the numerical quadrature with second-order accuracy is adopted for the integral. Secondly, terms related to the third derivative by taking $|\alpha|=3$ in $IS$ is ignored, and only terms related to the first and second derivatives are included. In smooth regions, the obtained $IS$ approximates the analytical one as
\begin{equation}\label{IS-accuracy}
IS=\sum_{i=1}^{3} A_i(W_ih)^2(1+O(h)),
\end{equation}
where $A_i$ are the parameters dependent on cell's geometry, and $W_i$ are three first derivatives. The degradation of $IS$ accuracy will not affect the accuracy of final nonlinear reconstruction.

\subsection{WENO weight with improved adaptivity}
Adaptive variation of the magnitude of $\tau_Z$ in WENO weight is important for the accuracy and robustness of the WENO scheme. $\tau_Z$ is required to adaptively satisfy $\tau_Z/IS_k=O(h^r) (r>0)$ for the candidate polynomials for smooth solutions, and it satisfies $\tau_Z/IS_k=O(1)$ for the candidates crossing discontinuities.
Thus the definition of $\tau_Z$ is directly related to the accuracy and robustness of the scheme.
For the current high-order simplified WENO reconstruction (accuracy order is higher than 2), the orders of the candidate polynomials include zero, one and two or three.
As a result, the separation of order of the candidate polynomials results in a large deviation in the values of smoothness indicators.
For example, the fourth-order nonlinear reconstruction is obtained from $p^3$, $p^1$ and $p^0$.
A small first derivative of $p^1$ or large second and third derivatives of $p^3$ will lead to large deviations in the values of their $IS_k$ even in smooth region, and a small first derivative of $p^1$ is common at critical point.

In this paper, $\tau_Z$ is defined by the hierarchical differences of $IS_k$. $\tau_Z$ consists of two parts. The first part is the difference of two $IS_k$ which are more likely to have small deviation in values in smooth region.
To further identify discontinuity, the second part is obtained by two $IS_k$ whose values are more likely to have large deviation.
$\tau_Z$ can be uniformly written as
\begin{equation}\label{tau-z}
\tau_Z=\tau_Z^{HO} +(\tau_Z^{LO})^{\beta},
\end{equation}
where $\beta$ is an adaptive power.
For the fourth-order reconstruction, $\tau_Z$ can be given as
\begin{equation}\label{tau-z-4th}
\tau_Z=\big| IS_0-IS^* \big| +\big| IS^*-IS^{**} \big|^{\beta}.
\end{equation}
$IS_0$ is from the candidate polynomial $p_0$, $IS^*$ is the auxiliary one with smaller deviation from $IS_0$ in values, and $IS^{**}$ is the auxiliary one with larger deviation with $IS^{*}$ in the vicinity of discontinuities.
$IS^*$ is given by Eq.(\ref{IS}) based on $p^2$ which is obtained from the third-order compact reconstruction.
$IS^{**}$ is taken as $IS^{**}=(\sum_{k=1}^4 IS_k-\mathrm{max}_{1\leq k \leq4}{IS_k}-\mathrm{min}_{1\leq k \leq4}{IS_k}$)/2 which corresponds to smooth candidate polynomials (assuming only one discontinuity exists at four faces of the reconstructed cell) but not the smoothest one whose $IS_k$ may be close to $0$.
The first term of $\tau_Z$ in Eq.(\ref{tau-z-4th}) is going to be a higher-order small value, even at critical points. While the second term of $\tau_Z$ in Eq.(\ref{tau-z-4th}) can be $O(1)$ in the vicinity of discontinuities.
For the third-order and second-order reconstructions, the same $\tau_Z$ as in Eq.(\ref{tau-z-4th}) is used, but different $IS^*$ is defined.
In this case, $IS^*$ is taken as $IS^*=\mathrm{max}_{1\leq k\leq 4}\{IS_k\}$.
$\beta$ is defined as
\begin{equation}\label{pow-q}
\beta =\left\{\begin{array}{ll}
2,  & C_0 \big|\frac{IS^{**}+\epsilon}{IS^*+\epsilon}-1 \big| <C_1 h, \\
1,  & \mathrm{otherwise},
\end{array} \right.
\end{equation}
where the free-parameters are uniformly taken as $C_1=1$ and $C_0=\mathrm{min}\{\mathrm{sgn}(\rho_{thres}-\rho_0)+1,\mathrm{sgn}(p_{thres}-p_0)+1,1\}$. $\rho_0$ and $p_0$ are the minimum cell-averaged density and pressure in the reconstructed cell and its first-level neighboring cells. $\rho_{thres}$ and $p_{thres}$ are uniformly taken as $\rho_{thres}=p_{thres}=5.0\times 10^{-2}$ in this paper.

The improved adaptivity of $\tau_Z$ is achieved by the hierarchical difference of indicators and the adaptive power $\beta$.
The high-order part $\tau_Z^{HO}$ of $\tau_Z$ can be a high-order small value in smooth region, and the lower-order part $\tau_Z^{LO}$ of $\tau_Z$ can be reduced to a high-order small value in smooth region by taking $\beta=2$. At the same time, discontinuities of variables on the stencil can be identified in the current reconstruction sensitively. The first part and the second part will take values of order $O(1)$ in discontinuities, which makes the nonlinear weights of smoother  polynomials candidates have values $O(1)$.
In particular, $C_0$ in the definition of $\beta$ improves the robustness of the scheme in the regions with a very small value of density or pressure.
.

Recall that the values of $IS_s$ and $IS_d$ in Section 3 are taken as $IS_d=IS_0$ and $IS_s=IS^{**}$.

\section{Numerical examples}
In this section, a few test cases on 3D tetrahedral meshes will be conducted.
The time step is determined by the CFL condition.
In all test cases, CFL number is taken as $CFL\geq0.5$, except CFL$=0.35$ in the Taylor-Green vortex flow.
For viscous flows, the time step is also limited by the viscous term as $\Delta t=CFL h^2/(3\nu)$,
where $h$ is the cell size and $\nu$ is the kinematic viscosity coefficient.
The collision time $\tau$ for inviscid flow at a cell interface is defined by
\begin{align*}
\tau=\varepsilon \Delta t + \varepsilon_{\mathrm{diss}}\displaystyle|\frac{p_l-p_r}{p_l+p_r}|\Delta t,
\end{align*}
where $\varepsilon=0.05$, $\varepsilon_{\mathrm{diss}}=10$, and $p_l$ and $p_r$ are the pressures at the left and right sides of a cell interface.
For the viscous flow, the collision time is related to the viscosity coefficient,
\begin{align*}
\tau=\frac{\mu}{p} + \varepsilon_{\mathrm{diss}} \displaystyle|\frac{p_l-p_r}{p_l+p_r}|\Delta t,
\end{align*}
where $\mu$ is the dynamic viscosity coefficient and $p$ is the pressure at the cell interface.
In smooth flow region, it will reduce to $\tau=\mu/p$.
The reason for including pressure jump term in the particle collision time is to increase the shock wave thickness numerically to the order
of cell size \cite{xu2}.

\begin{figure}[!htb]
\centering
\includegraphics[width=0.50\textwidth]{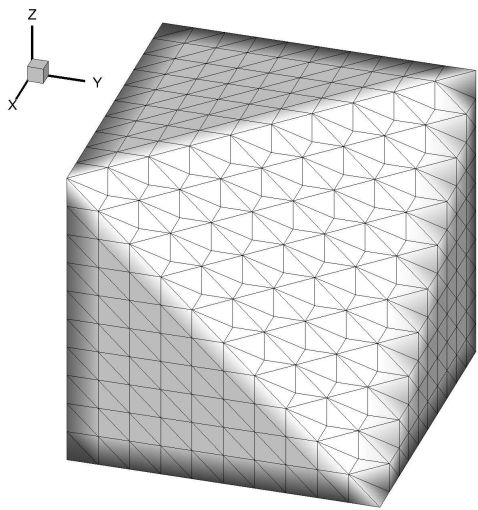}
\caption{\label{Mesh-accuracy} Accuracy test: the regular tetrahedral mesh with $10^3\times 6$ mesh DOFs. The tetrahedral mesh is obtained by dividing each regular hexahedral cell in the Cartesian mesh into six tetrahedral cells.}
\end{figure}

\subsection{Accuracy test}
The three-dimensional advection of density perturbation is used to verify the order of accuracy of compact GKS.
The initial condition is given by
\begin{align*}
\begin{split}
\rho(x,y,z)=1+0.2\sin(\pi (x+y+z)),~p(x,y,z)=1, \\
U(x,y,z)=V(x,y,z)=W(x,y,z)=1.
\end{split}
\end{align*}
The computational domain is $[0,2]^3$. The periodic boundary conditions are applied on all domain boundaries.
The tetrahedral mesh is used in the computation. The tetrahedral mesh is obtained by dividing each regular hexahedral cell in the Cartesian mesh into six tetrahedral cells. The coarsest mesh used in the computation is shown in Fig. \ref{Mesh-accuracy}.
The $L^1$ and $L^{\infty}$ errors and convergence orders obtained by the compact GKS with linear reconstruction at $t=2$ are presented in Table \ref{Accu-sin-linear}.
Due to the non-uniform mesh cell, the accuracy order of $L^{\infty}$ cannot reflect the true convergence order of the scheme.
From the numerical results listed in Table \ref{Accu-sin-linear}, it can be seen that the accuracy orders of $L^1$ are almost the same as the theoretical ones.

\begin{figure}[!htb]
\centering
\includegraphics[width=0.495\textwidth]{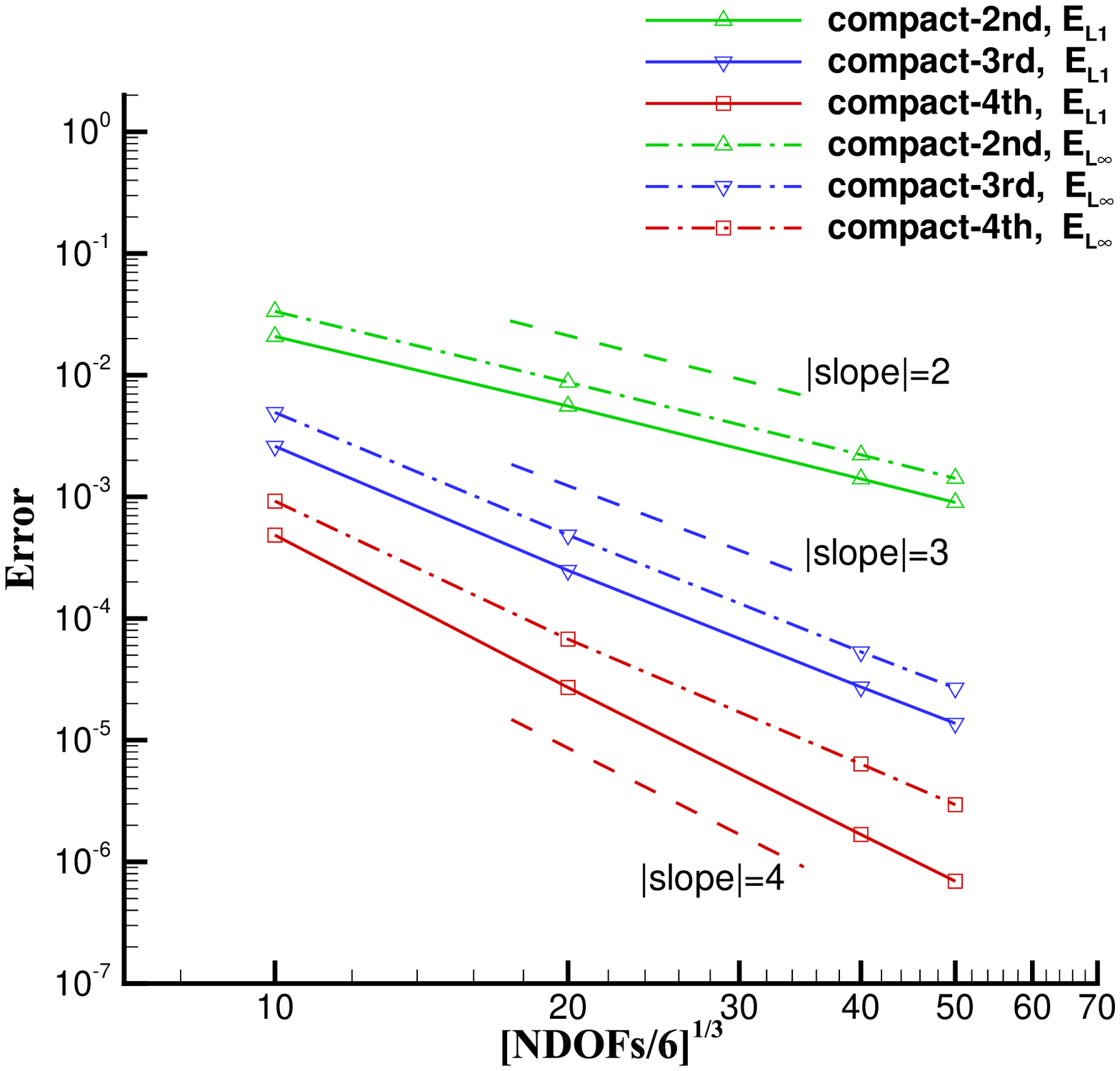}
\includegraphics[width=0.495\textwidth]{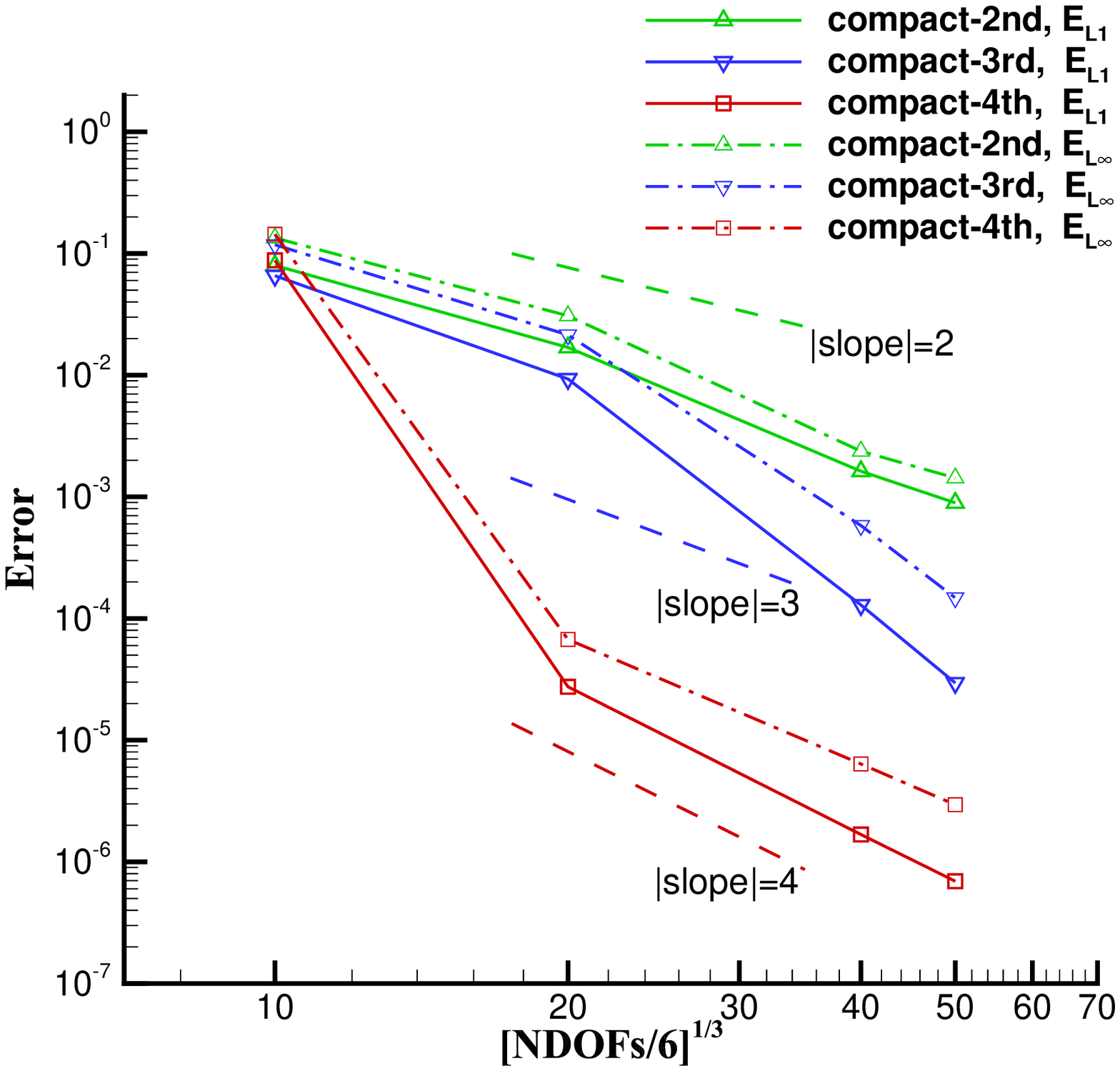}
\caption{\label{Accuracy-dof} Accuracy test: errors versus mesh DOFs in one direction by compact GKS with linear reconstruction (left) and nonlinear reconstruction (right).}
\end{figure}

The errors versus mesh DOFs is given in Fig. \ref{Accuracy-dof} for linear and nonlinear compact GKS,
where the abscissa represents the mesh DOFs in one direction.
Compared with the second-order scheme, the high-order scheme has a smaller error under the same mesh DOFs; at the same time, under the same error requirement, the mesh DOF used by the high-order scheme is far less than second-order one.
For example, when the error limit is $10^{-3}$, the required mesh DOFs by the linear fourth-order scheme are nearly two orders of magnitude less than that by the linear second-order scheme, and the required mesh DOFs by the nonlinear fourth-order scheme is $1/35$ of that by the nonlinear second-order scheme.

\begin{figure}[!htb]
\centering
\includegraphics[width=0.495\textwidth]{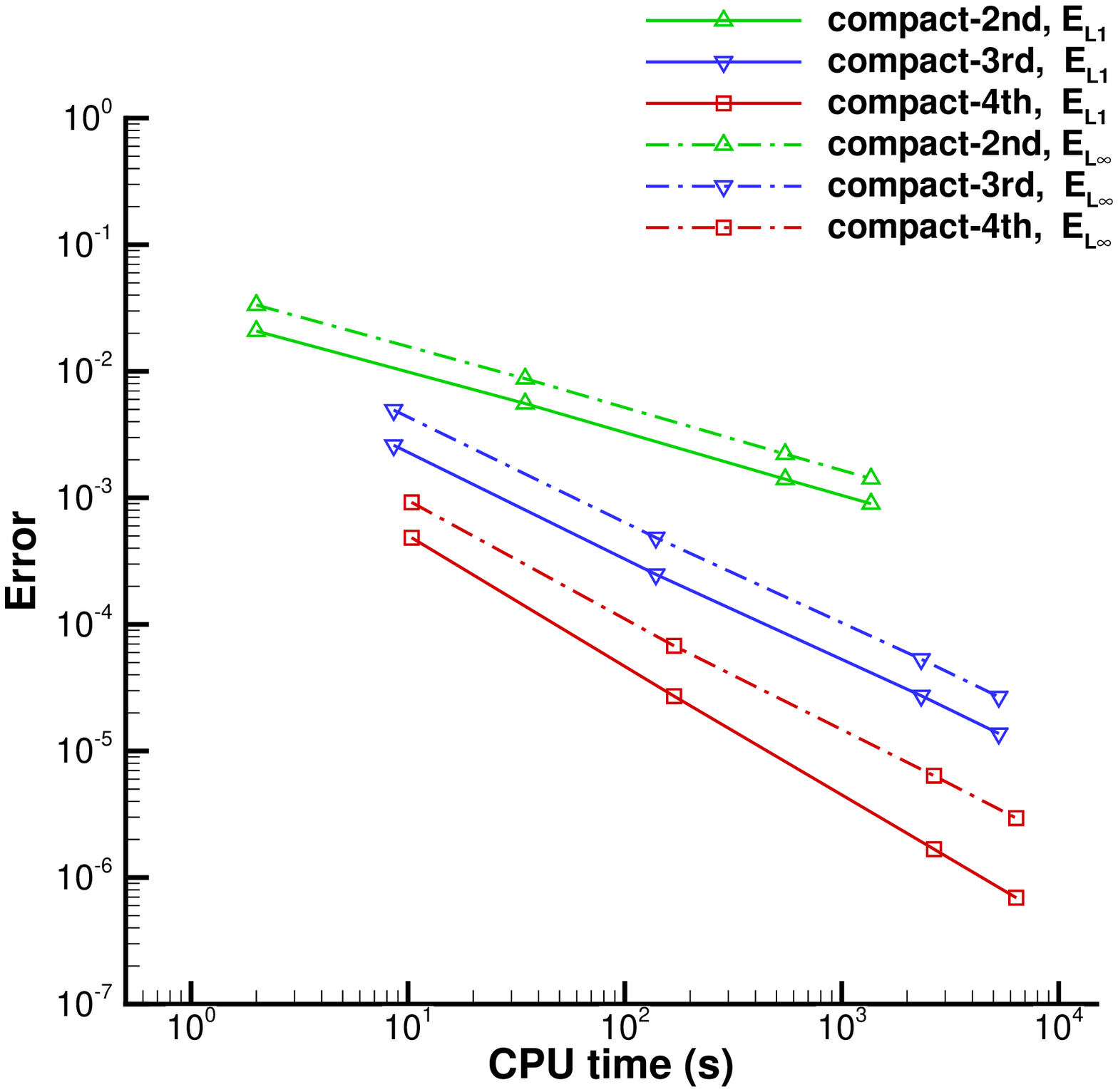}
\includegraphics[width=0.495\textwidth]{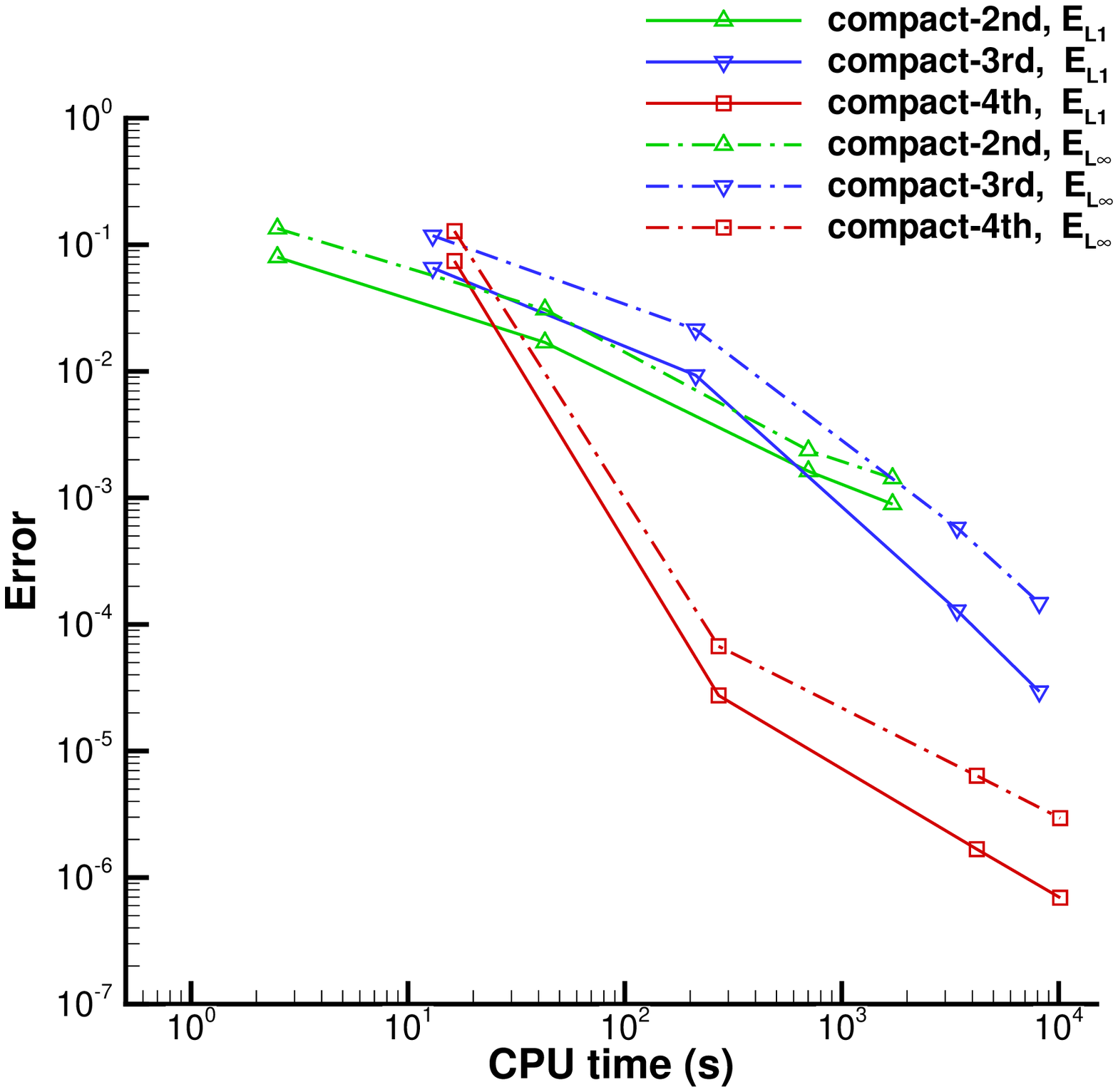}
\caption{\label{Accuracy-time} Accuracy test: errors versus CPU time by compact GKS with linear reconstruction (left) and nonlinear reconstruction (right).}
\end{figure}

Fig. \ref{Accuracy-time} lists the relationship between error and CPU time. The computation is performed by an OpenMP parallel code using $48$ threads on a $2.2$ GHz Intel(R) Xeon(R) workstation.
The curves of errors versus CPU time are basically similar to the curves of errors versus mesh DOFs of one direction.
The results indicate that the increase in CPU time relevant to the increase in algorithm complexity of high-order compact scheme accounts for a small proportion in the overall CPU time in comparison with the second-order scheme.

\begin{table}[!h]
	\small
	\begin{center}
		\def\temptablewidth{1.0\textwidth}
		{\rule{\temptablewidth}{0.50pt}}
        \footnotesize
		\begin{tabular*}{\temptablewidth}{@{\extracolsep{\fill}}c c cc cc}
            scheme      & $h_{re}$ &$Error_{L^1}$ &$\mathcal{O}_{L^1}$ &$Error_{L^{\infty}}$ &$\mathcal{O}_{L^{\infty}}$ \\
			\hline
            2nd-order   
            compact GKS & $10^3\times 6$  & 2.0826e-02    & ~     & 3.3425e-02  & ~      \\
                        & $20^3\times 6$  & 5.5685e-03    & 1.90  & 8.7536e-03  & 1.93   \\
                        & $40^3\times 6$  & 1.4054e-03    & 1.99  & 2.2178e-03  & 1.98   \\
                        & $50^3\times 6$  & 8.9934e-04    & 2.00  & 2.2178e-03  & 2.00   \\

			\hline
            3rd-order   
            compact GKS & $10^3\times 6$  & 2.6059e-03    & ~     & 4.9418e-03  & ~      \\
                        & $20^3\times 6$  & 2.4874e-04    & 3.39  & 4.8530e-04  & 3.35   \\
                        & $40^3\times 6$  & 2.7360e-05    & 3.18  & 5.3194e-05  & 3.19   \\
                        & $50^3\times 6$  & 1.3751e-05    & 3.08  & 2.6821e-05  & 3.07   \\

			\hline
            4th-order   
            compact GKS & $10^3\times 6$  & 4.8439e-04    & ~     & 9.2211e-04  & ~      \\
                        & $20^3\times 6$  & 2.7104e-05    & 4.16  & 6.7724e-05  & 3.77   \\
                        & $40^3\times 6$  & 1.6770e-06    & 4.01  & 6.3761e-06  & 3.41   \\
                        & $50^3\times 6$  & 6.9290e-07    & 3.96  & 2.9498e-06  & 3.45   \\
		\end{tabular*}
		{\rule{\temptablewidth}{0.50pt}}
	\end{center}
	\vspace{-6mm} \caption{\label{Accu-sin-linear} Accuracy test: errors and convergence orders at $t=2$ obtained by linear compact GKS on regular tetrahedral meshes.}
\end{table}

\begin{figure}[!htb]
\centering
\includegraphics[width=0.495\textwidth]{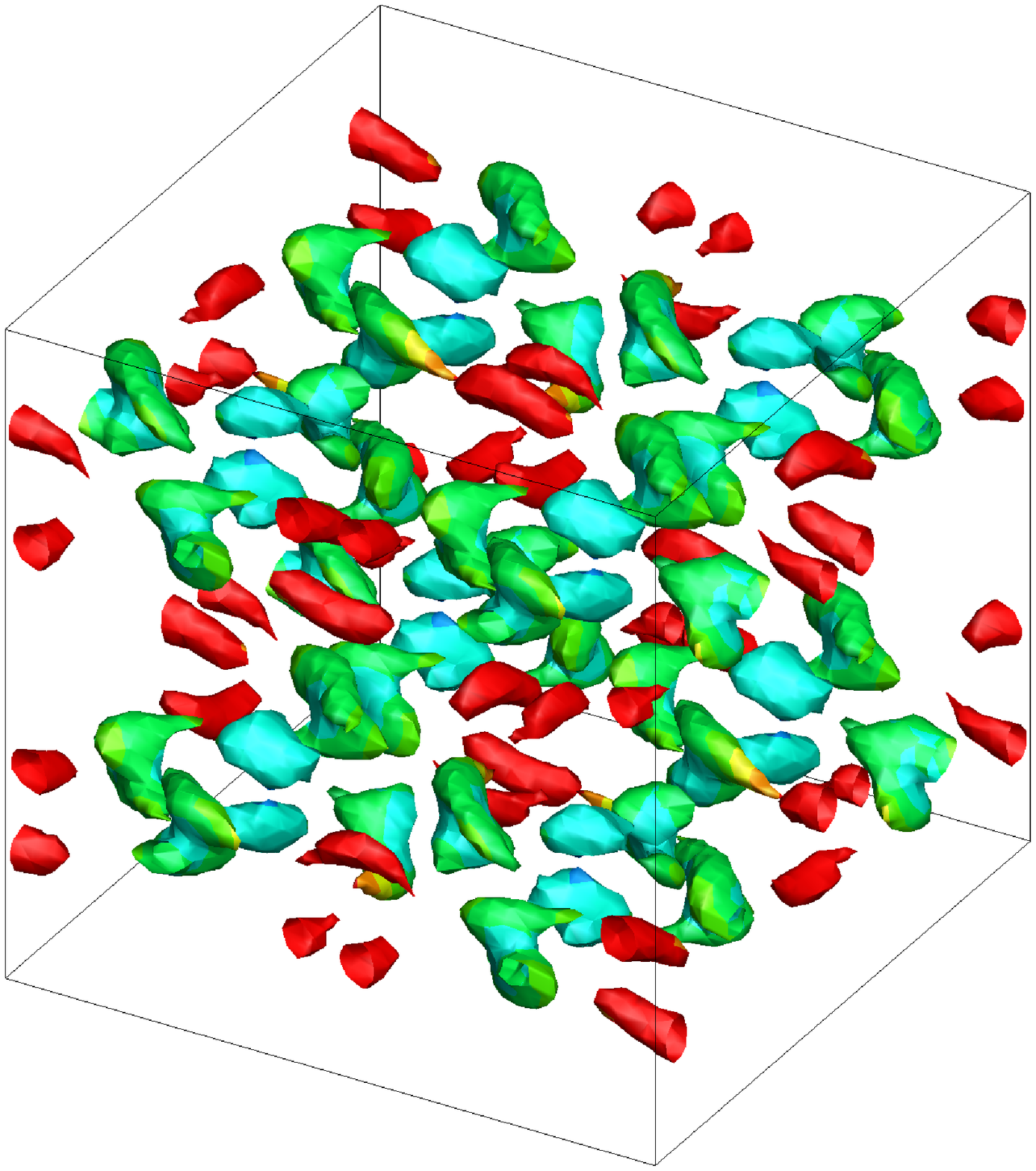}
\includegraphics[width=0.495\textwidth]{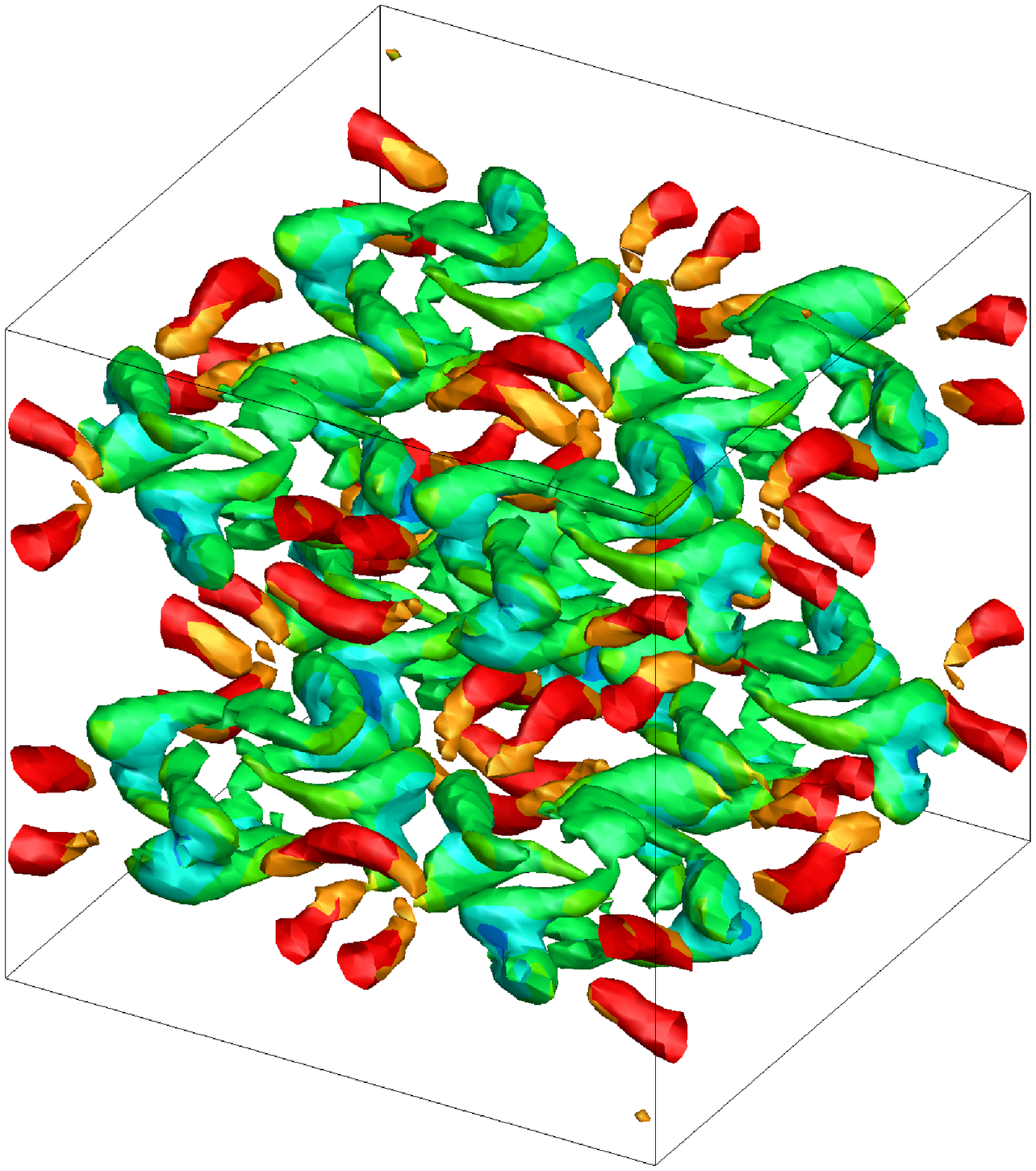}
\caption{\label{tgv-1} Taylor-Green vortex flow at $Re=280$: iso-surface of the second invariant of velocity gradient tensor $Qv=0.2$ colored by pressure at $t=15$. The results are obtained by second-order (left) and fourth-order (right) compact GKS. The number of mesh cells is $40^3\times6$.}
\end{figure}

\begin{figure}[!htb]
\centering
\includegraphics[width=0.495\textwidth]{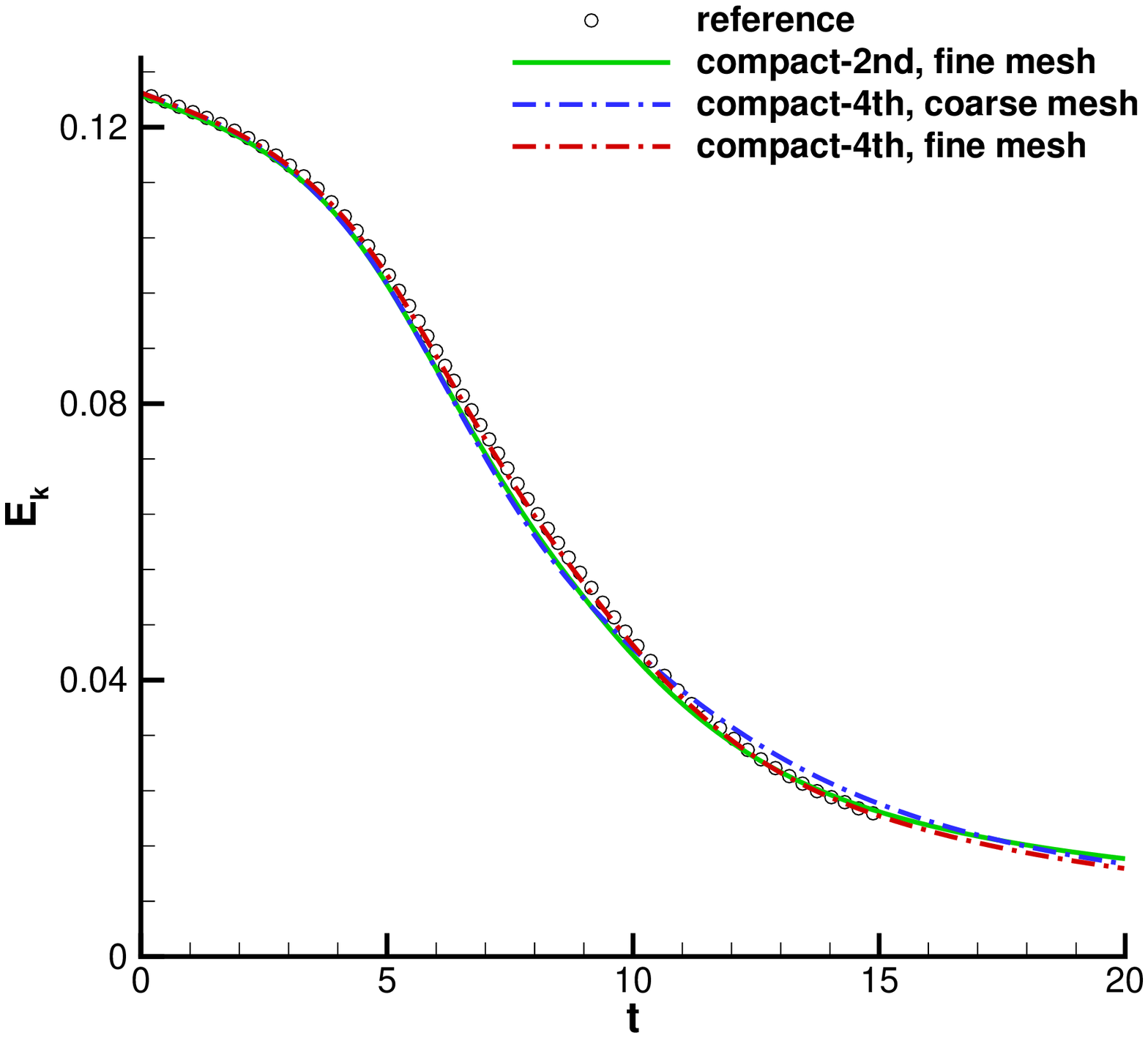}
\includegraphics[width=0.495\textwidth]{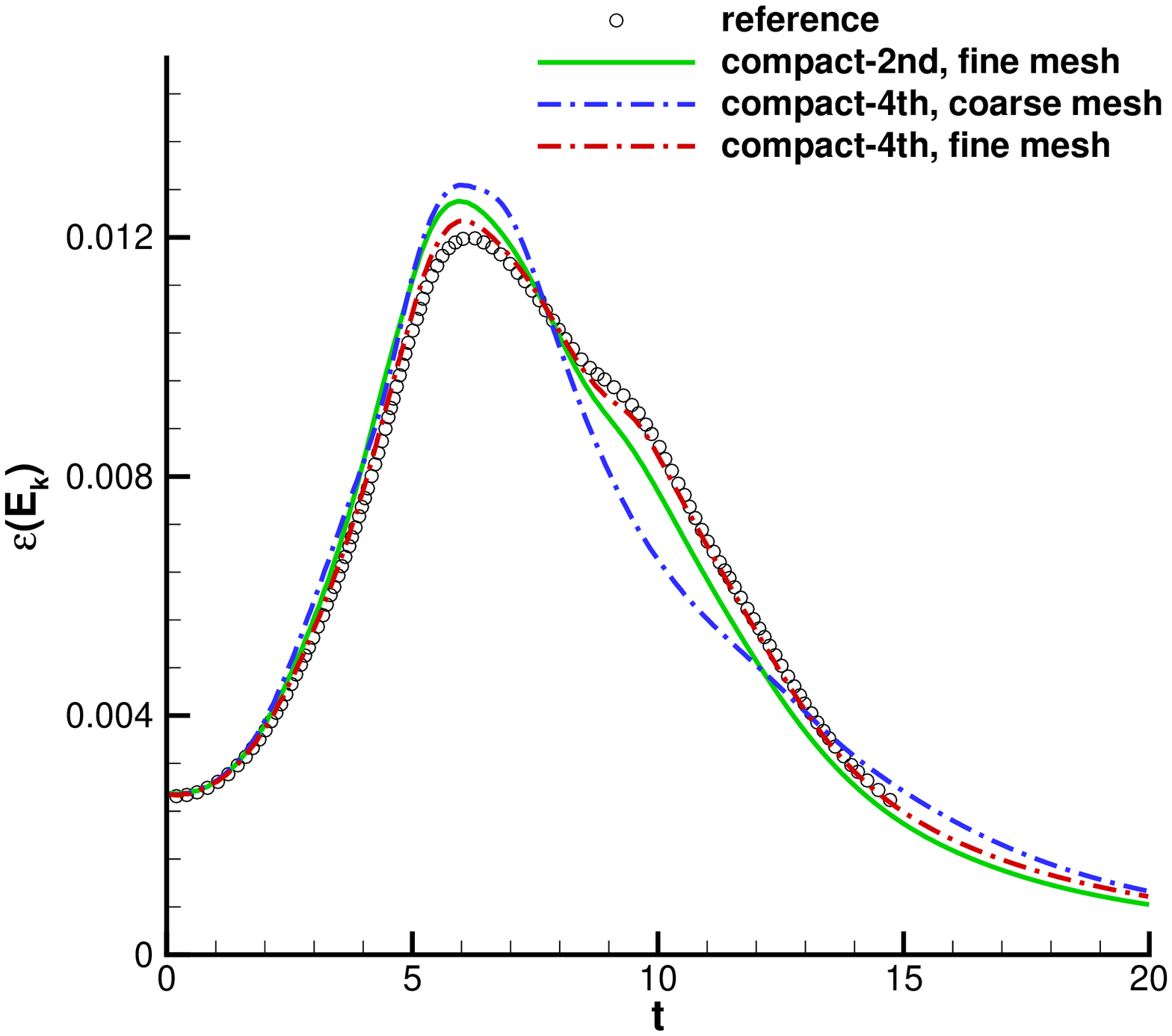}
\caption{\label{tgv-2} Taylor-Green vortex flow at $Re=280$: time history of kinetic energy and its dissipation rate. The numbers of mesh cells of the coarse mesh and fine mesh are $20^3\times6$ and $40^3\times6$ respectively.}
\end{figure}

\subsection{Taylor-Green vortex flow}
Taylor-Green vortex flow is a popular test to assess high-order schemes \cite{tgv_re280-1,tgv_re280-2}, which is used to verify the accuracy of the schemes for viscous flow.
Starting from a smooth initial flow distribution,  small scale flow structure in the flow field will emerge and evolve continuously.
The initial condition is set as
\begin{equation*}
\begin{split}
&U= U_0\mathrm{sin}x \mathrm{cos}y \mathrm{cos}z,\\
&V=-U_0\mathrm{cos}x \mathrm{sin}y \mathrm{cos}z,\\
&W=0,\\
&p=p_0+\frac{\rho_0 U_0^2}{16}(\mathrm{cos}2x+\mathrm{cos}2y)(\mathrm{cos}2z+2),
\end{split}
\end{equation*}
where $U_0=1$, $\rho_0=1$. The Mach number is $Ma=0.1$ and it is determined by $Ma=U_0/\sqrt{\gamma p_0/\rho_0}$.
The Reynolds number is $Re=280$ defined by $Re=\rho_0 U_0/\mu$,
where $\mu$ is the dynamic viscosity coefficient. The computational domain is $[-\pi,\pi]^3$.
This test case is used to verify the accuracy and the linear stability of the current compact GKS,
and only the linear compact GKS is adopted in the computation.
In order to compare the performance of second-order and fourth-order compact GKS, a coarse mesh with $20^3\times 6$ cells and a fine mesh with $40^3\times 6$ cells are used in the computation, with the same tetrahedral mesh used in the previous accuracy test.
The averaged kinetic energy is defined as
\begin{align*}
E_k=\frac{1}{|\Omega|}\int_{\Omega} \frac{1}{2}\mathbf{U}\cdot \mathbf{U}\mathrm{d}V,
\end{align*}
where $\Omega$ is the computational domain. $E_k$ is calculated by numerical quadrature. \
For the fourth-order scheme, the five-point Gaussian quadrature formula is used,
where the values of flow variables at the quadrature points are obtained by the same fourth-order compact reconstruction.
For second-order scheme, the mid-point quadrature formula is adopted.
The dissipation rate of kinetic energy is given by
\begin{align*}
\varepsilon(E_k)=\frac{\mathrm{d} E_k}{\mathrm{d} t}.
\end{align*}
$\varepsilon(E_k)$ is calculated by central difference method with second-order accuracy by considering the small difference in the time steps.

Fig. \ref{tgv-1} presents the iso-surface of the second invariant of velocity gradient tensor $Qv=0.18$ at $t=15$ colored by pressure,
where the left and right figures are the results from second-order and fourth-order compact schemes, respectively.
The fourth-order compact GKS has better resolution than that from the second-order scheme.
The time history of $E_k$ and $\varepsilon(E_k)$ are shown in Fig. \ref{tgv-1}. The reference solution is from \cite{tgv_re280-1}.
The result of the second-order compact GKS on fine mesh is slightly better than that of the fourth-order compact GKS on coarse mesh. The results of the fourth-order compact GKS on fine mesh has a good agreement with the reference solution.

\begin{figure}[!htb]
\centering
\includegraphics[width=0.45\textwidth]{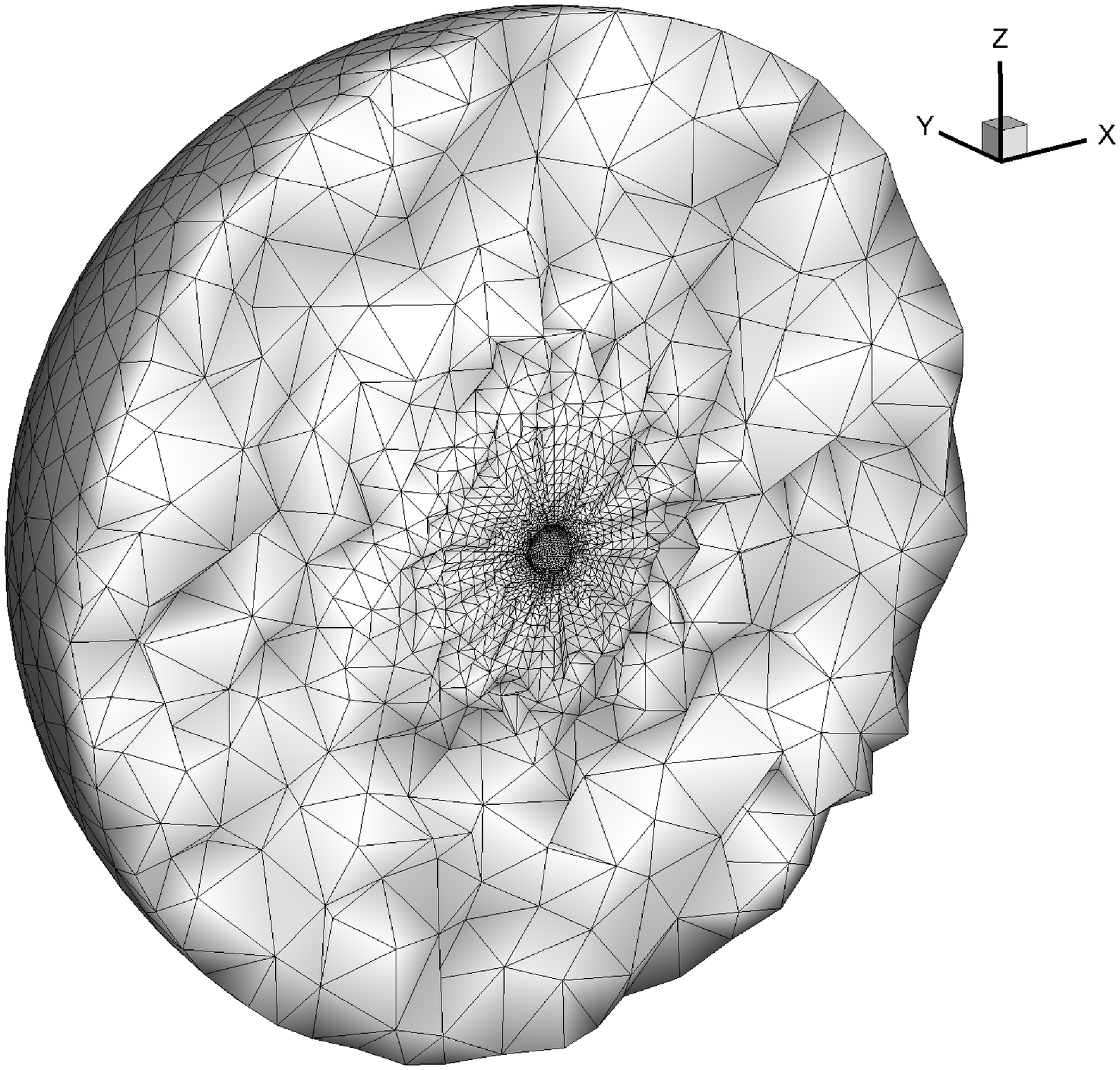}
\includegraphics[width=0.45\textwidth]{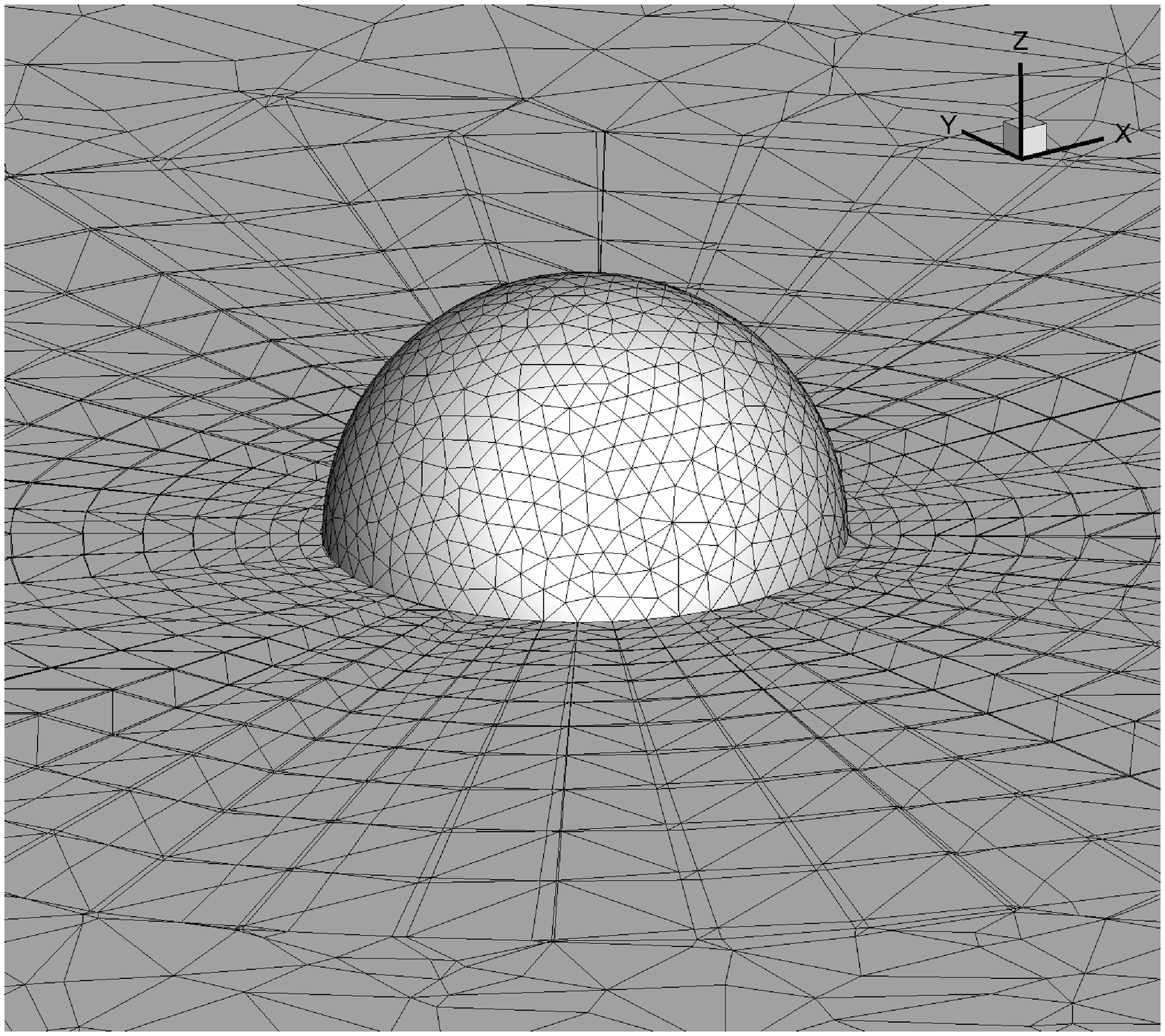}
\caption{\label{2-vis-sphere-1} Subsonic viscous flow around a sphere: the left is the mesh distribution over half of the computational domain, and the right is a local enlargement of the mesh around the sphere with diameter $D=1$. The outer boundary of the computational domain is a spherical surface with radius $10D$. }
\end{figure}

\begin{figure}[!htb]
\centering
\includegraphics[width=0.45\textwidth]{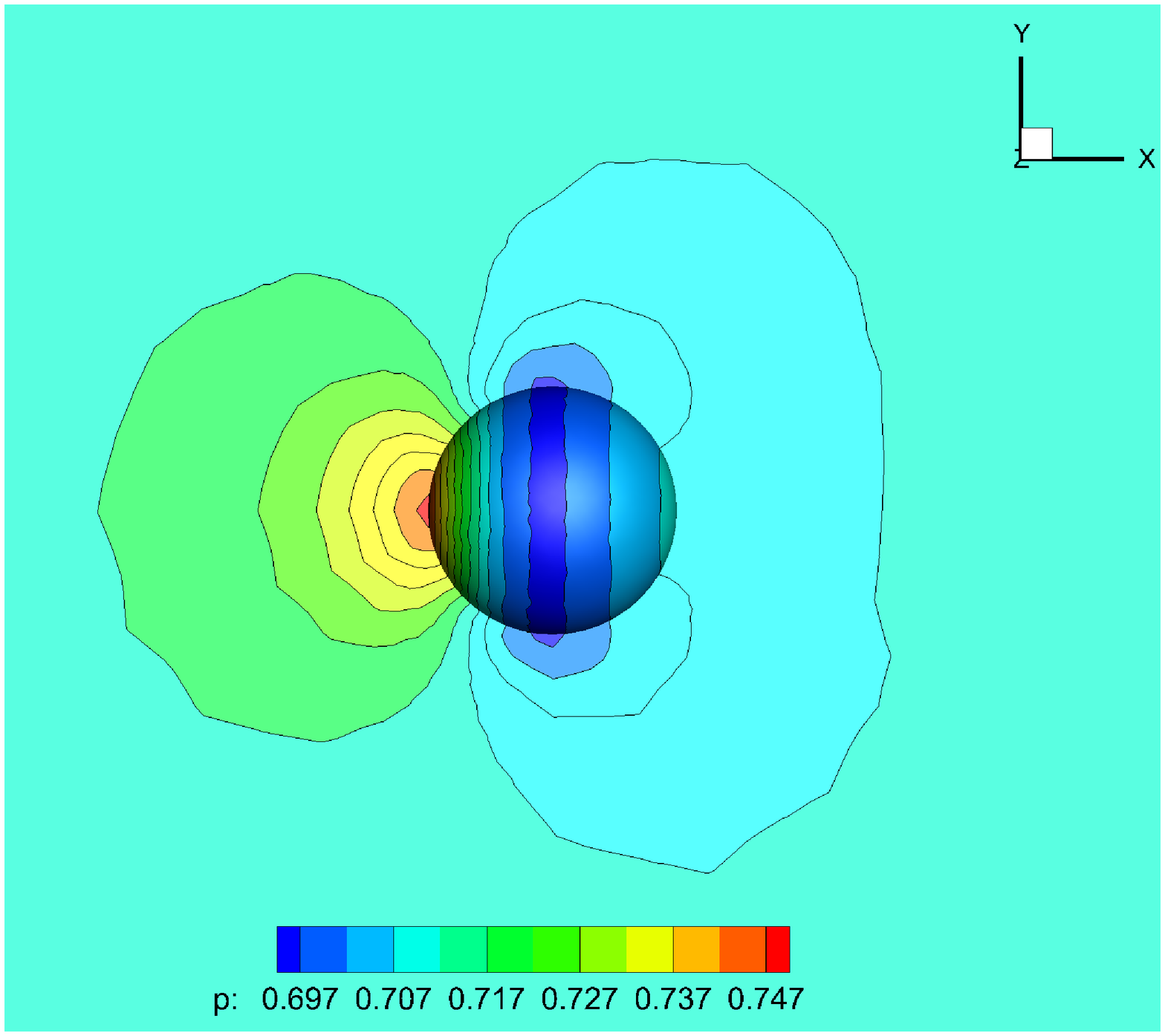}
\includegraphics[width=0.45\textwidth]{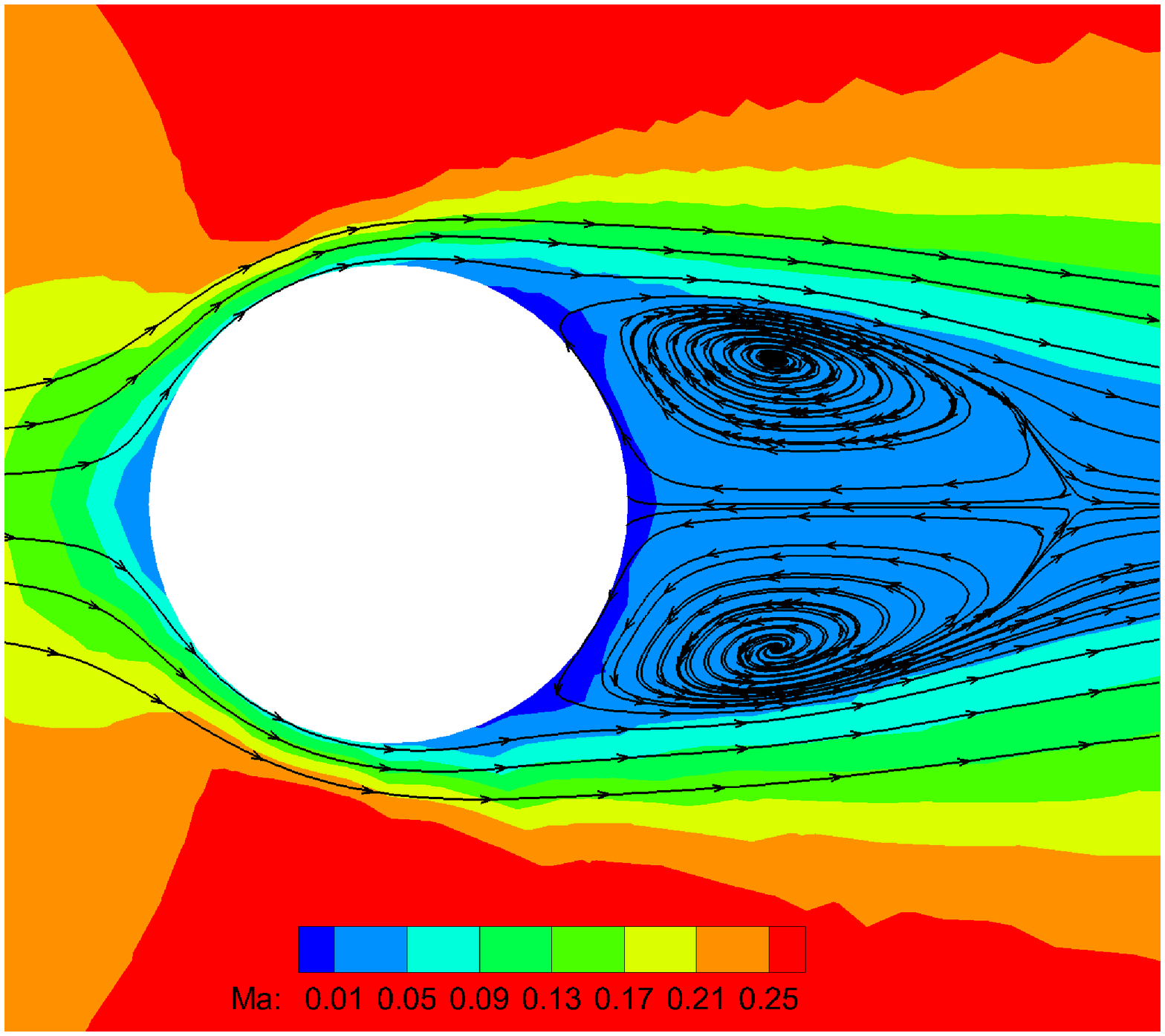}
\caption{\label{2-vis-sphere-2} Subsonic viscous flow around a sphere: the 3-D pressure contours (left) and 2-D streamlines on $z=0$ plane (right) obtained by linear fourth-order compact GKS. The Mach number and Reynolds number of the incoming flow are $Re=118$ and $Ma=0.2535$.}
\end{figure}

\begin{table}[!h]
	\small
	\begin{center}
		\def\temptablewidth{1.0\textwidth}
		{\rule{\temptablewidth}{1.0pt}}
		\footnotesize
		\begin{tabular*}{\temptablewidth}{@{\extracolsep{\fill}}c|c|c|c|c}
			Scheme 		    				& computational  mesh         		                                &$C_D$  &$C_L$ &$L$   \\
			\hline
            Linear compact GKS-4th                                    & $116308$ cells, tetrahedral mesh    &1.0163 &$<3.0\times10^{-5}$ &0.941  \\
            WENO compact GKS-4th                                      & $116308$ cells, tetrahedral mesh    &1.0146 &$<1.5\times10^{-3}$ &0.907  \\
            VFV-4th with AMR \cite{sphere-re118-vfv}                  & $621440$ cells, hexahedral mesh    	&1.0157 &--                  &--     \\
			Direct DG scheme \cite{sphere-re118-dg}                   & $160868$ cells, mixed mesh          &1.0162 &--                  &0.96   \\
            $5$th-order hybrid scheme(FR and DG)\cite{sphere-re118-fr}& $68510$  cells, mixed mesh    	    &1.0162 &--                  &--     \\
		\end{tabular*}
		{\rule{\temptablewidth}{1.0pt}}
	\end{center}
	\vspace{-6mm} \caption{\label{2-vis-sphere-3} Subsonic viscous flow around a sphere: drag coefficient $C_D$ and lift coefficient $C_L$, and wake length $L$ obtained by different schemes. The Mach number and Reynolds number of the incoming flow are $Re=118$ and $Ma=0.2535$.}
\end{table}

\subsection{Subsonic viscous flow around a sphere}
The subsonic flow around a sphere is used to test the compact GKS in the capturing viscous flow solution.
The incoming flow is a uniform flow with Mach number $Ma=0.2535$ and Reynolds number $Re=118$, where the Reynolds number is evaluated by the diameter $D=1$ of the sphere.
At this Reynolds number, the flow is steady and there are two attached vortices on the downwind side of the sphere.
The outer boundary of the computational domain is a spherical surface with radius $10D$. The tetrahedral mesh in Fig. \ref{2-vis-sphere-1} is used. The size of the first layer cells on the sphere is $h_{min}=5\times10^{-2}$, and the cell size at the ending point of the attached vortices is $0.2$. The total number of the cells in the computational domain is $116308$. The CFL number takes $CFL=0.6$ in the computation.

The 3-D pressure contours on the sphere and $z=0$ plane and the 2-D streamline distribution obtained by the linear fourth-order compact GKS are given in Fig. \ref{2-vis-sphere-2}.
In order to verify that the nonlinear scheme in this paper has similar accuracy to the linear scheme, the fourth-order compact GKS with WENO reconstruction is also used in this test case.
The quantitive results of the compact GKS and other computations are given in Table \ref{2-vis-sphere-3}, where the drag and lift coefficients, wake length $L$,  and the total number of mesh cells are listed for comparison. The linear fourth-order GKS gives the same value of $C_D$ as direct DG scheme \cite{sphere-re118-dg} and $5$th-order hybrid scheme of FR and DG \cite{sphere-re118-fr}, but uses fewer total DOFs, where $(r+1)(r+2)(r+3)/6$ DOFs are used in each cell for $(r+1)$th-order DG scheme. The nonlinear fourth-order GKS gives the $C_D$ with a $0.17\%$ error from the linear scheme, which validate the similar high-order accuracy of nonlinear fourth-order compact GKS to the linear one.

\begin{figure}[!htb]
\centering
\includegraphics[width=0.45\textwidth]{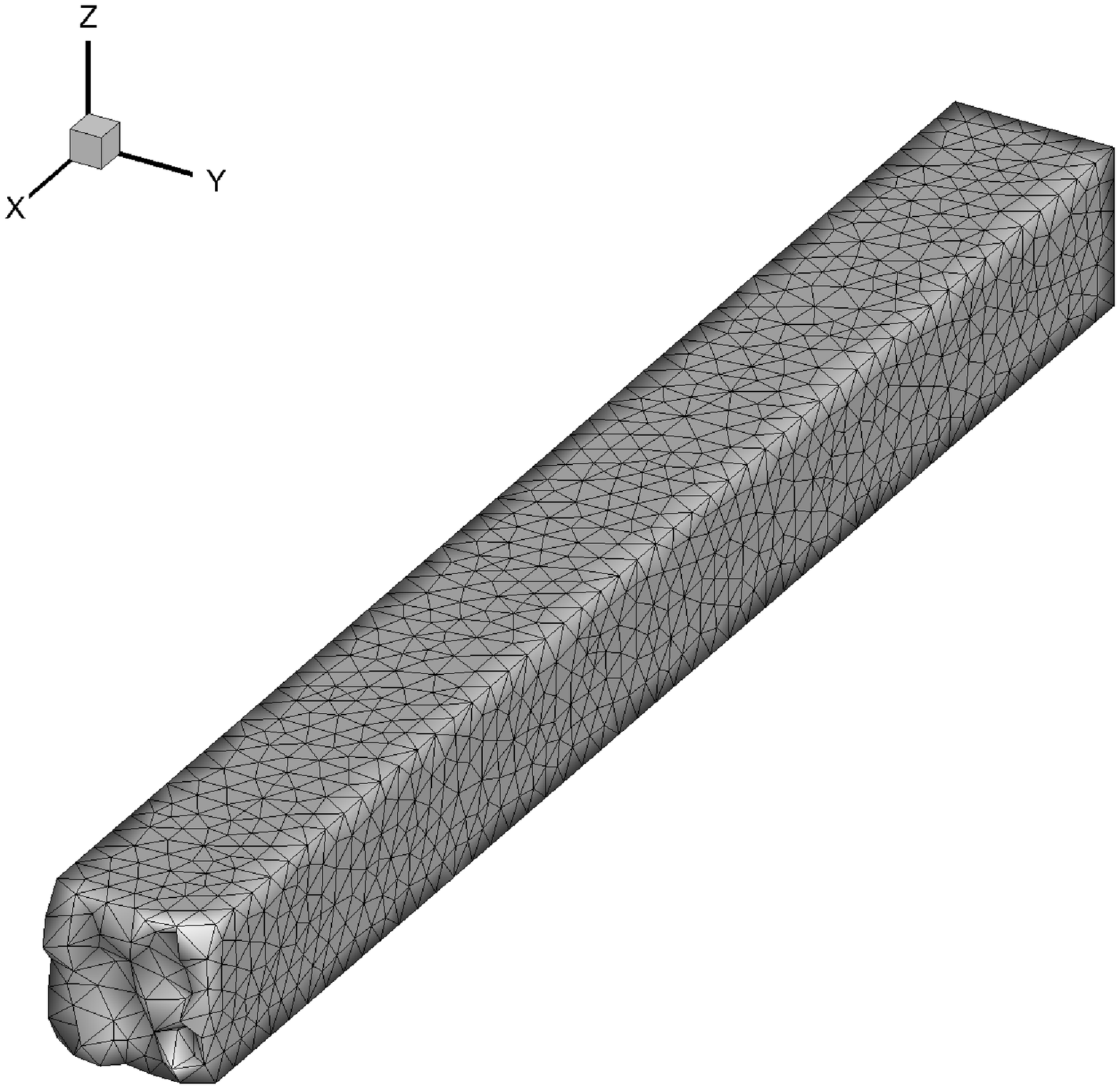}
\includegraphics[width=0.45\textwidth]{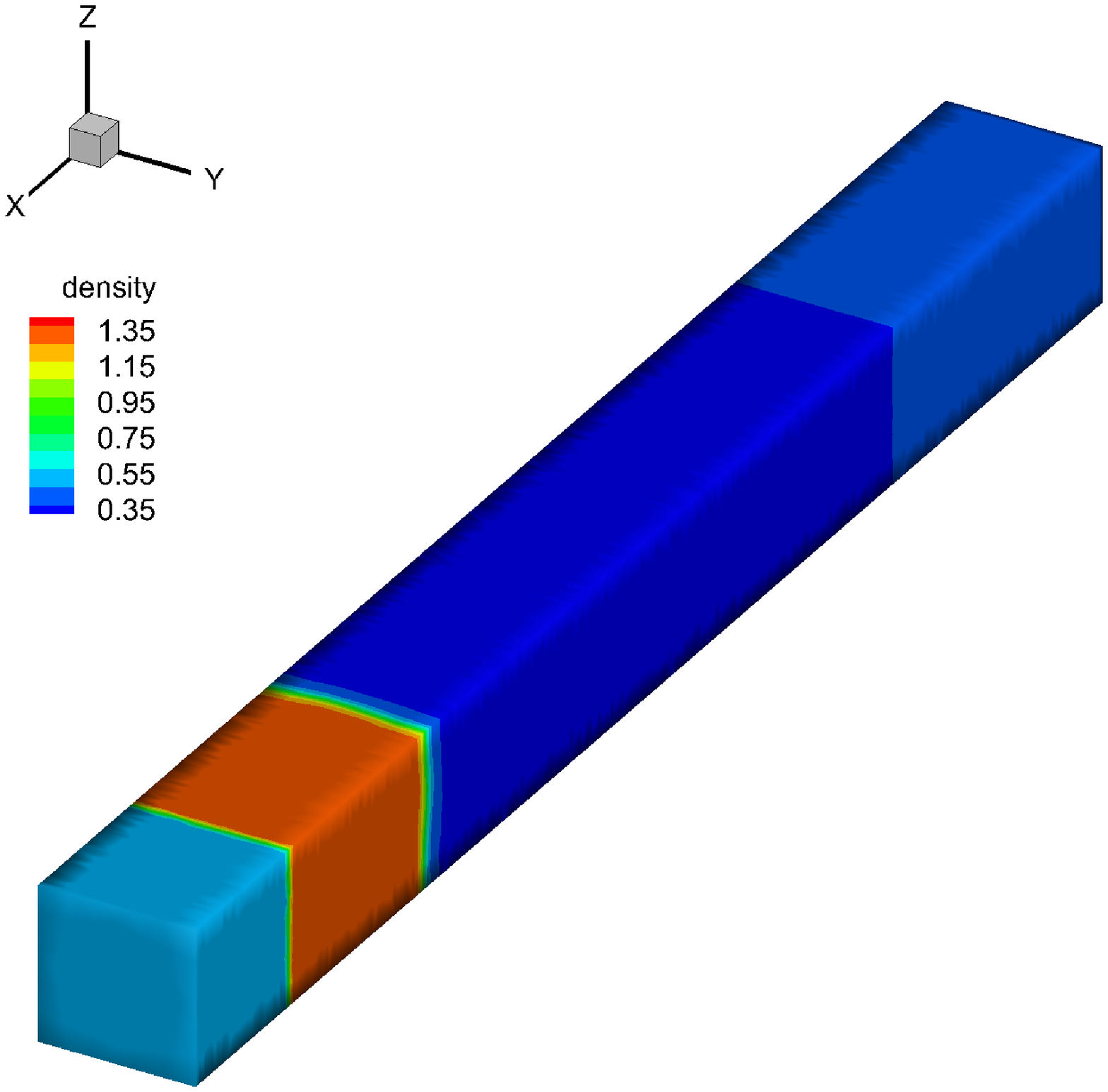}    \\
\includegraphics[width=0.45\textwidth]{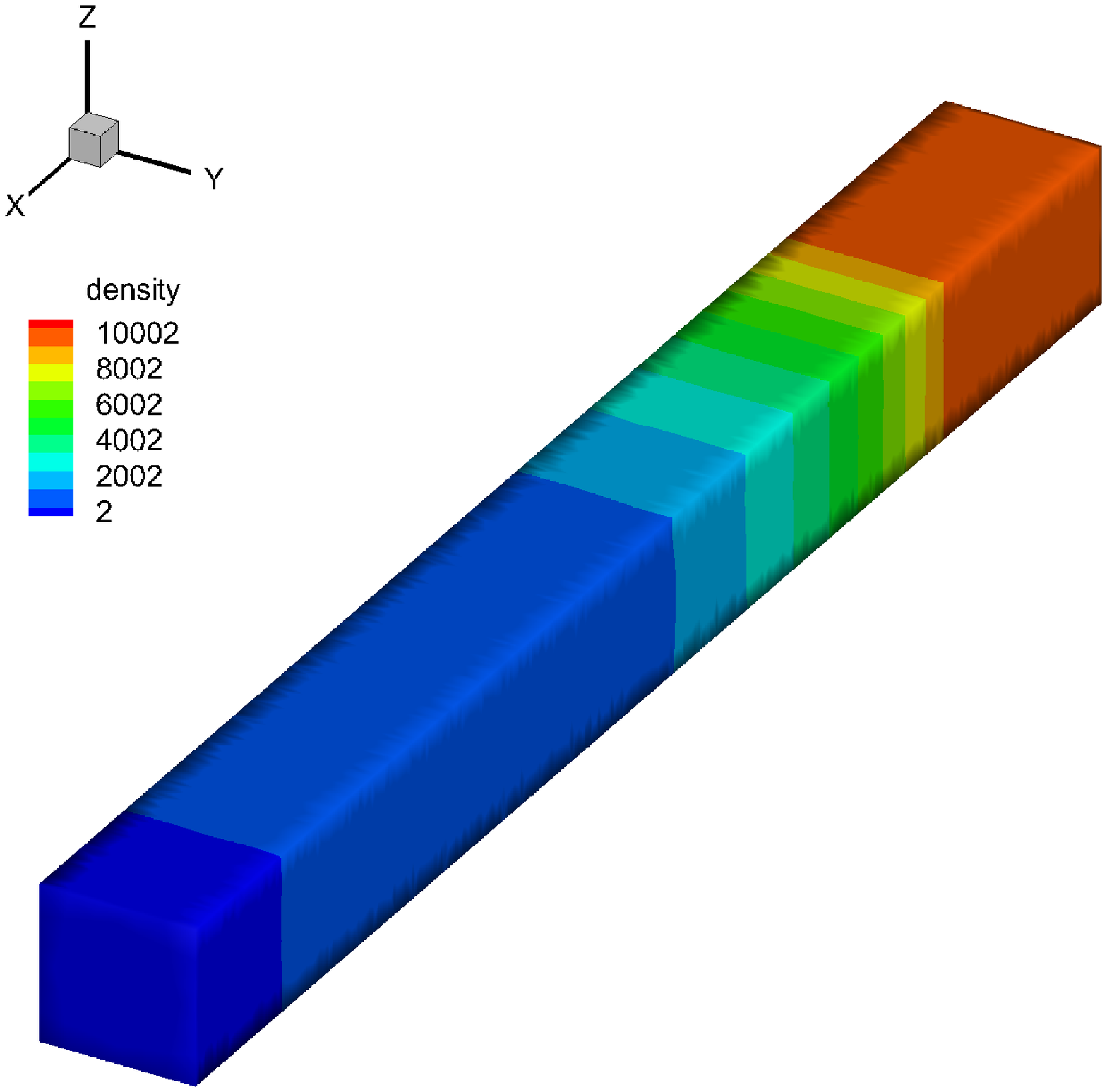}
\includegraphics[width=0.45\textwidth]{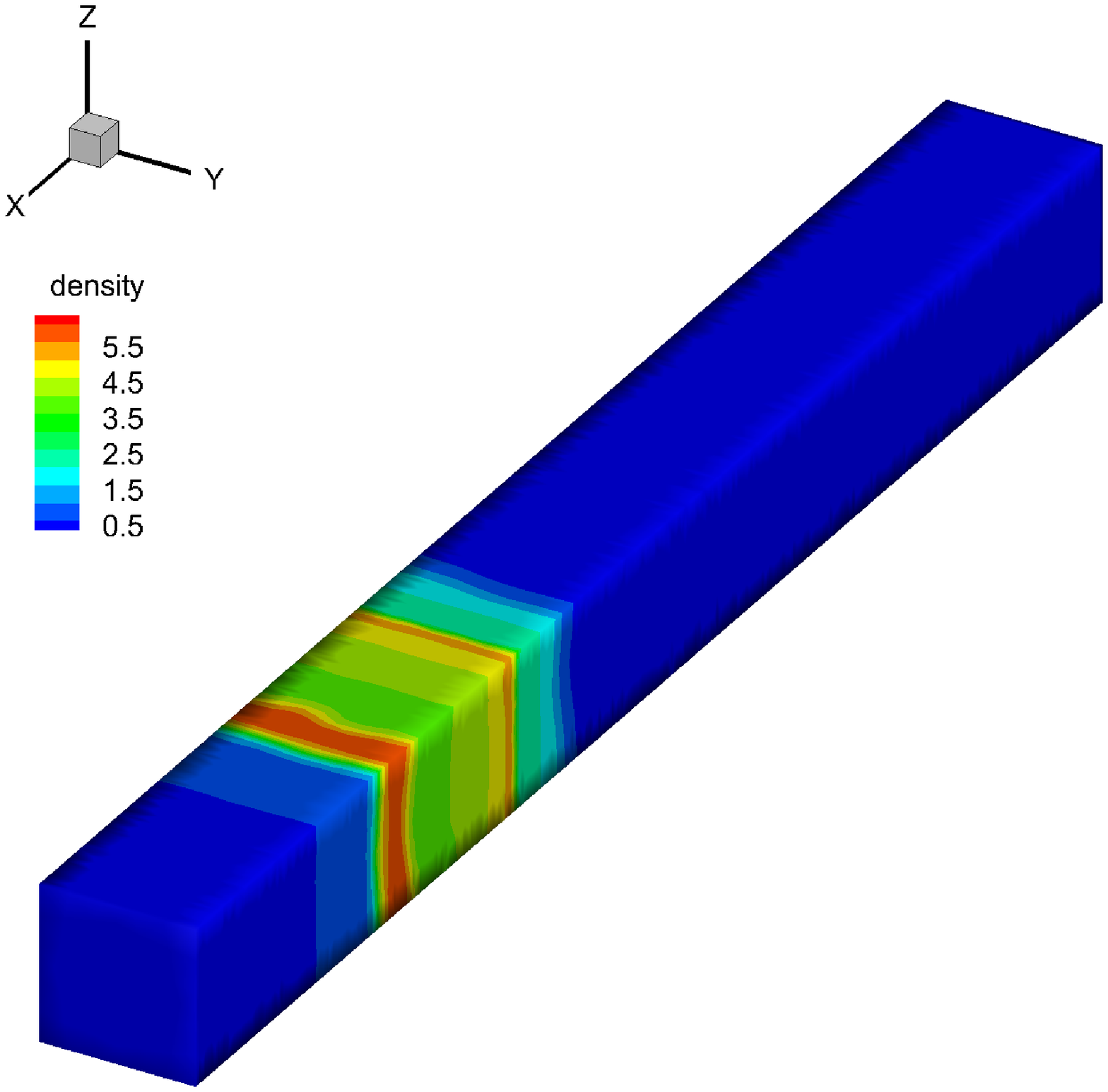}
\caption{\label{1-1d-3d} Riemann problems and blast wave problem: the local enlargement of the computational mesh and 3-D density contours of the three test cases obtained by the fourth-order compact GKS.}
\end{figure}

\begin{figure}[!htb]
\centering
\includegraphics[width=0.425\textwidth]{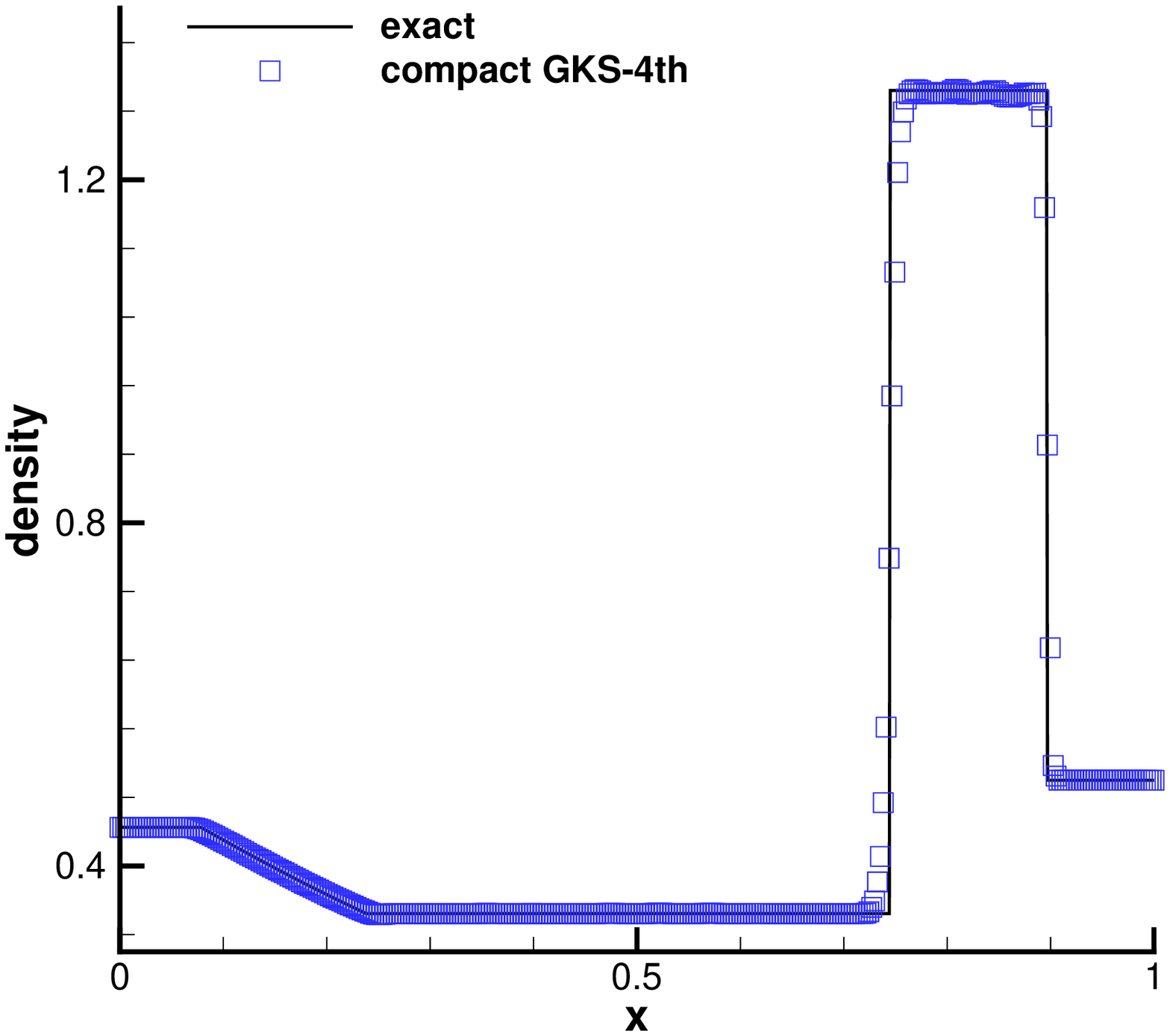}
\includegraphics[width=0.425\textwidth]{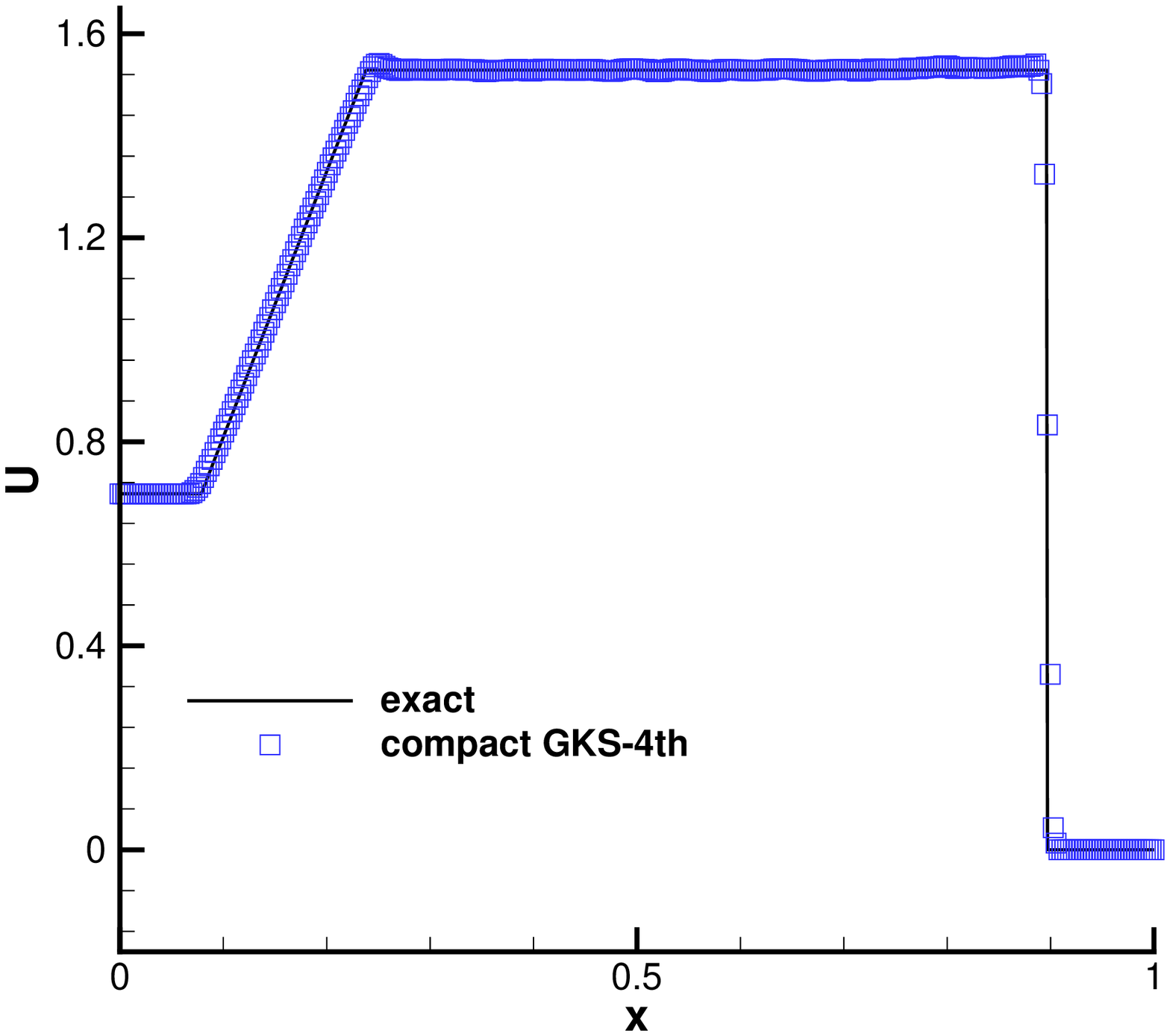}
\caption{\label{1-1d-1} Lax shock tube problem: numerical results and comparison with the exact solution at $t=0.16$ for density (left) and horizontal velocity (right). The 1-D result is the distribution along the x-direction centerline with $y=z=0.0125$.}
\end{figure}

\begin{figure}[!htb]
\centering
\includegraphics[width=0.425\textwidth]{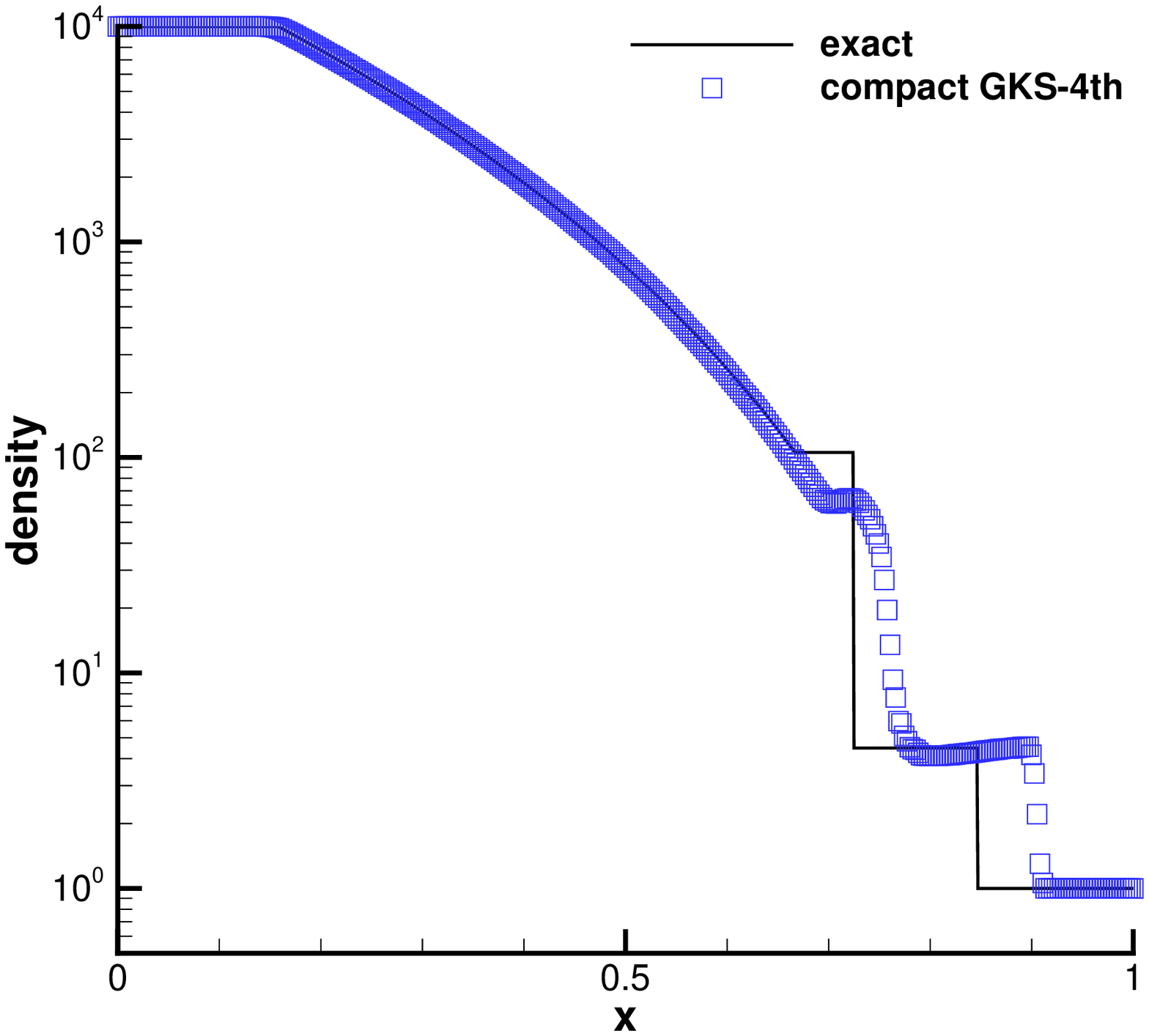}
\includegraphics[width=0.425\textwidth]{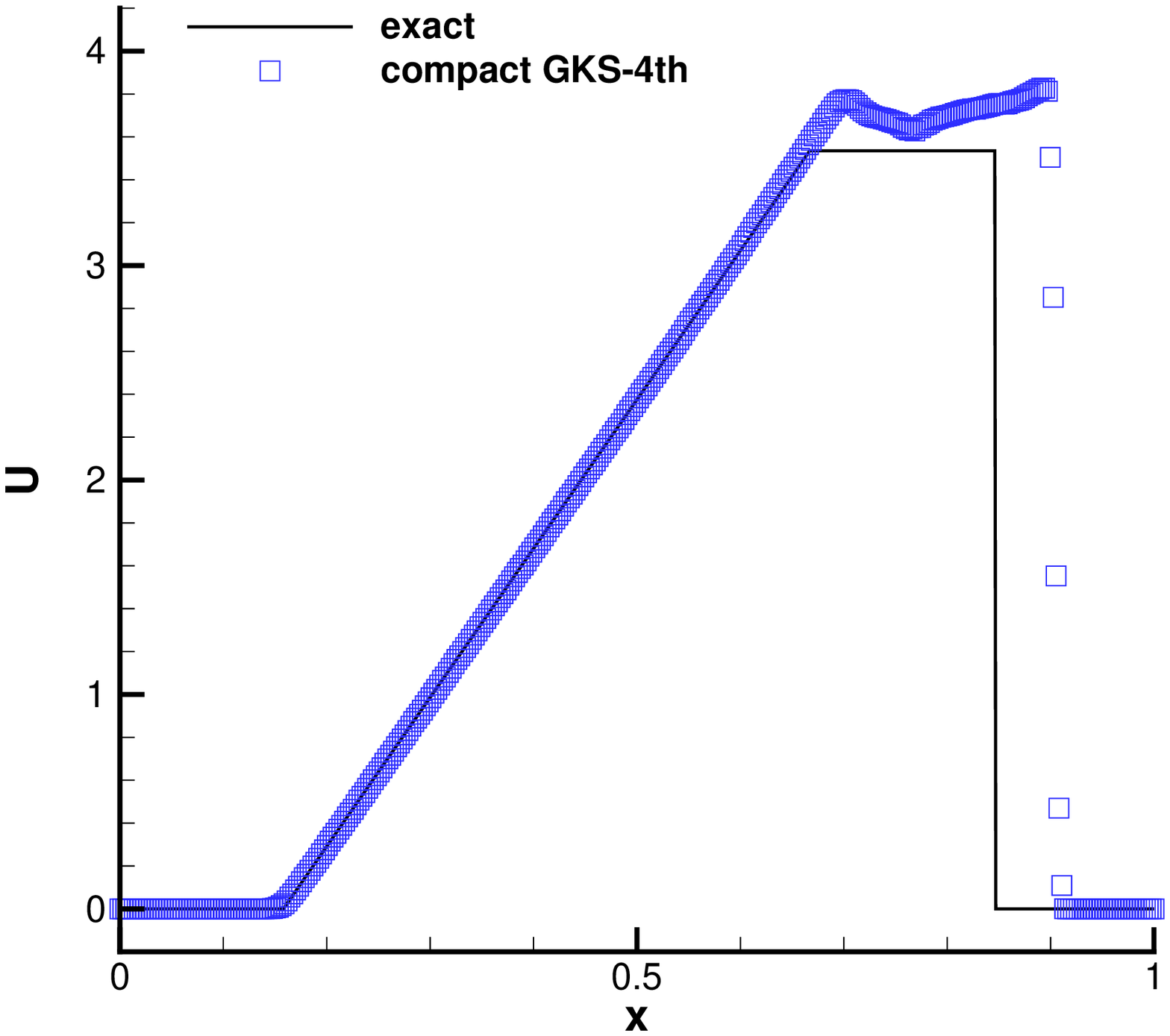}
\caption{\label{1-1d-2} Large pressure ratio problem: numerical results and comparison with the exact solution at $t=0.12$ for density (left) and horizontal velocity (right). The 1-D result is the distribution along the x-direction centerline with $y=z=0.0125$.}
\end{figure}

\begin{figure}[!htb]
\centering
\includegraphics[width=0.425\textwidth]{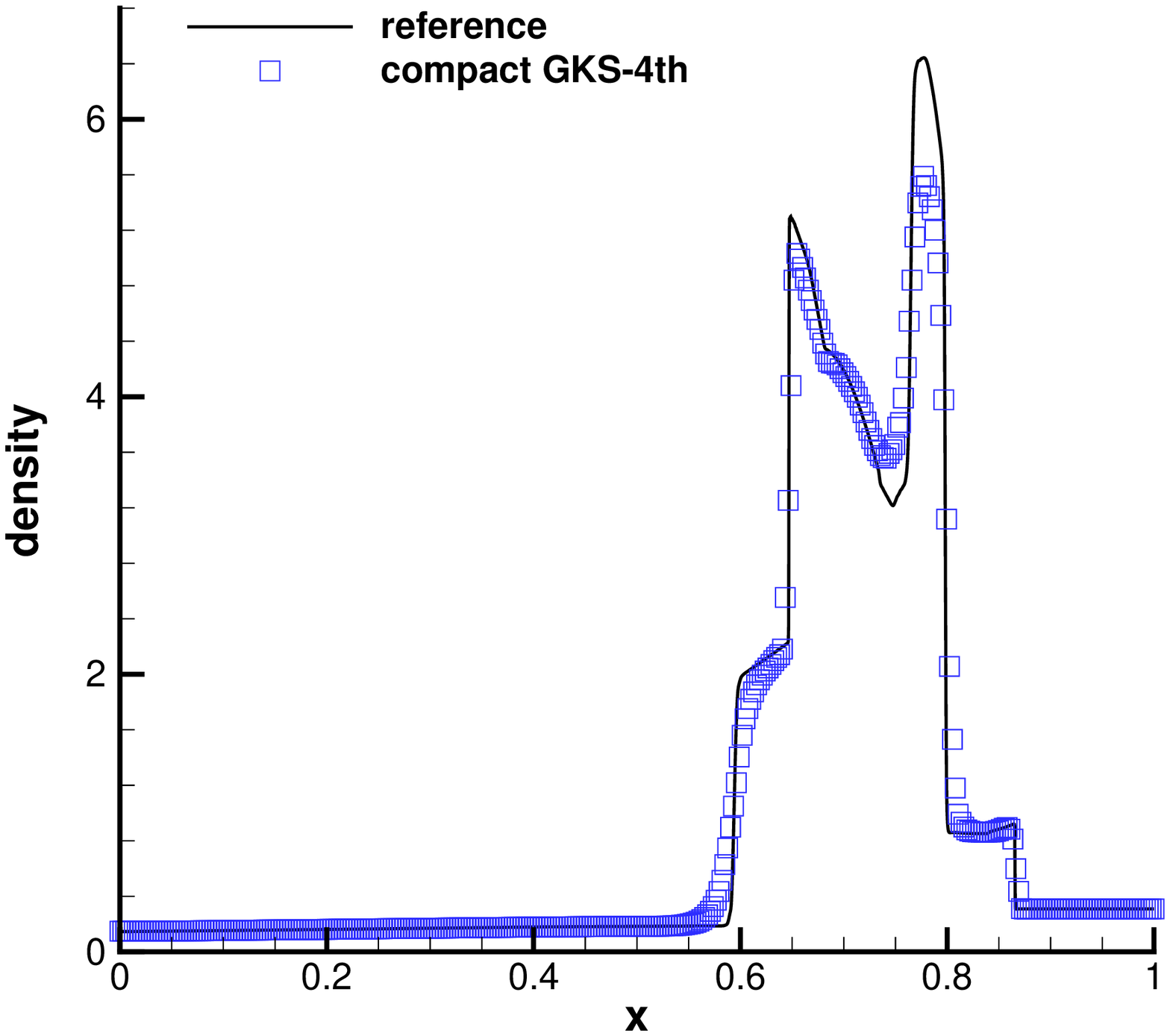}
\includegraphics[width=0.425\textwidth]{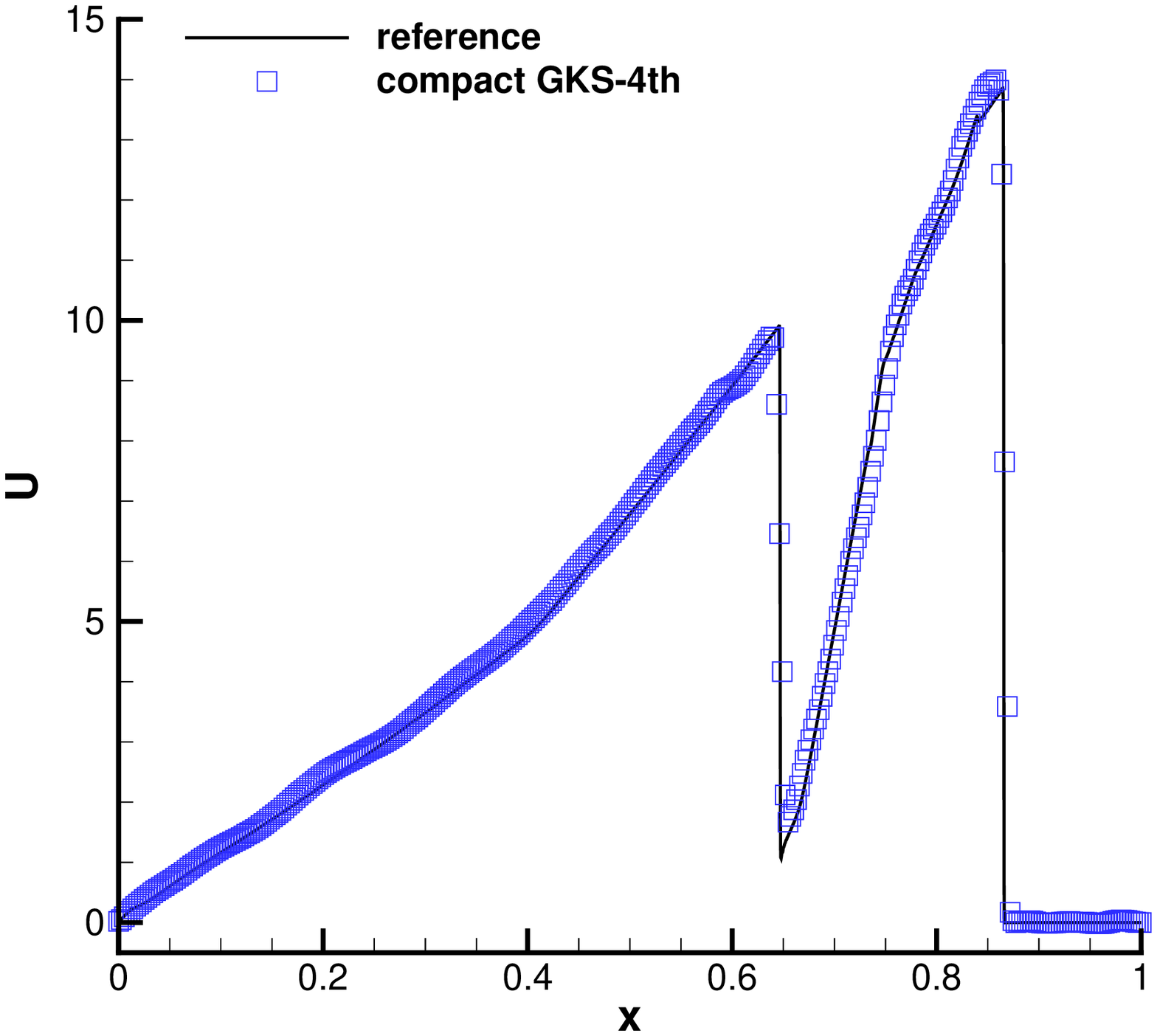}
\caption{\label{1-1d-3} Blast wave problem: numerical results and comparison with the reference solution at $t=0.038$ for density (left) and horizontal velocity (right). The 1-D result is the distribution along the x-direction centerline with $y=z=0.0125$.}
\end{figure}

\subsection{Riemann problems and blast wave problem}
To validate the robustness of the compact GKS, the scheme is applied to two 1-D Riemann problems and the blast wave problem for compressible inviscid flow.
The computational domain is given by $[0,1] \times [0,0.025]^2$. The tetrahedral mesh is used in the computation, and the average side length of tetrahedral cell is about $1/200$.
The initial conditions and the output times of two Riemann problems are listed in Table \ref{1d-RP}. For the Lax shock tube problem \cite{lax-testcase}, the fixed inflow condition and free boundary condition are used for the boundaries with $x=0$ and $x=1$ respectively, and the inviscid wall condition is imposed on the other boundaries. For the large pressure ratio problem \cite{tang-liu-testcase}, the free boundary condition is used on both ends at $x=0$ and $x=1$, and the inviscid wall condition is imposed on the other boundaries.
\begin{table}[!h]
	\small
	\begin{center}
		\def\temptablewidth{1.0\textwidth}
		{\rule{\temptablewidth}{1.0pt}}
		\footnotesize
		\begin{tabular*}{\temptablewidth}{@{\extracolsep{\fill}}c|c|c|c|c|c|c|c|c}
			test cases 		    		 & $\rho_L$         & $U_L$         & $p_L$   &$\rho_R$  & $U_R$ & $p_R$  &$x_D$   &$t_f$    \\
			\hline
			Lax shock tube problem       & 0.445          &0.698        & 3.528 &0.5     &0    &0.571 &0.5   &0.16    \\
			Large pressure ratio problem & $1\times10^4$    &0      & $1\times10^4$ &1       &0    &1     &0.3   &0.12    \\
		\end{tabular*}
		{\rule{\temptablewidth}{1.0pt}}
	\end{center}
	\vspace{-6mm} \caption{\label{1d-RP} Two 1-D Riemann problems for compressible inviscid flow. }
\end{table}

The initial condition of the blast wave problem \cite{woodward-testcase} is given as
\begin{align*}
(\rho,U,p) =\begin{cases}
(1, 0, 1000), & 0\leq x<0.1,\\
(1, 0, 0.01), & 0.1\leq x<0.9,\\
(1, 0, 100),  & 0.9\leq x\leq 1.
\end{cases}
\end{align*}
The output time is $t_f=0.038$.
The inviscid wall boundary condition is imposed on all boundaries for blast wave problem.

The local enlargement of the mesh and the 3-D contours of density are shown in Fig. \ref{1-1d-3d}.
The wave structures are accurately obtained by the fourth-order compact GKS on the tetrahedral meshes.
The computed density and velocity profiles of these cases are shown in Fig. \ref{1-1d-1}, Fig. \ref{1-1d-2} and Fig. \ref{1-1d-3}, respectively.
No numerical oscillations are observed near the shock fronts.
The compact GKS performs well on the large pressure jump and the blast wave problems.
On the current coarse mesh, especially tetrahedral one, the result of large pressure jump problem is reasonable in comparison with the results in \cite{tang-liu-testcase}, even the exact Riemann solver with shock fitting may not be helpful in this case on the tetrahedral mesh \cite{hui}.

\begin{figure}[!htb]
\centering
\includegraphics[width=0.425\textwidth]{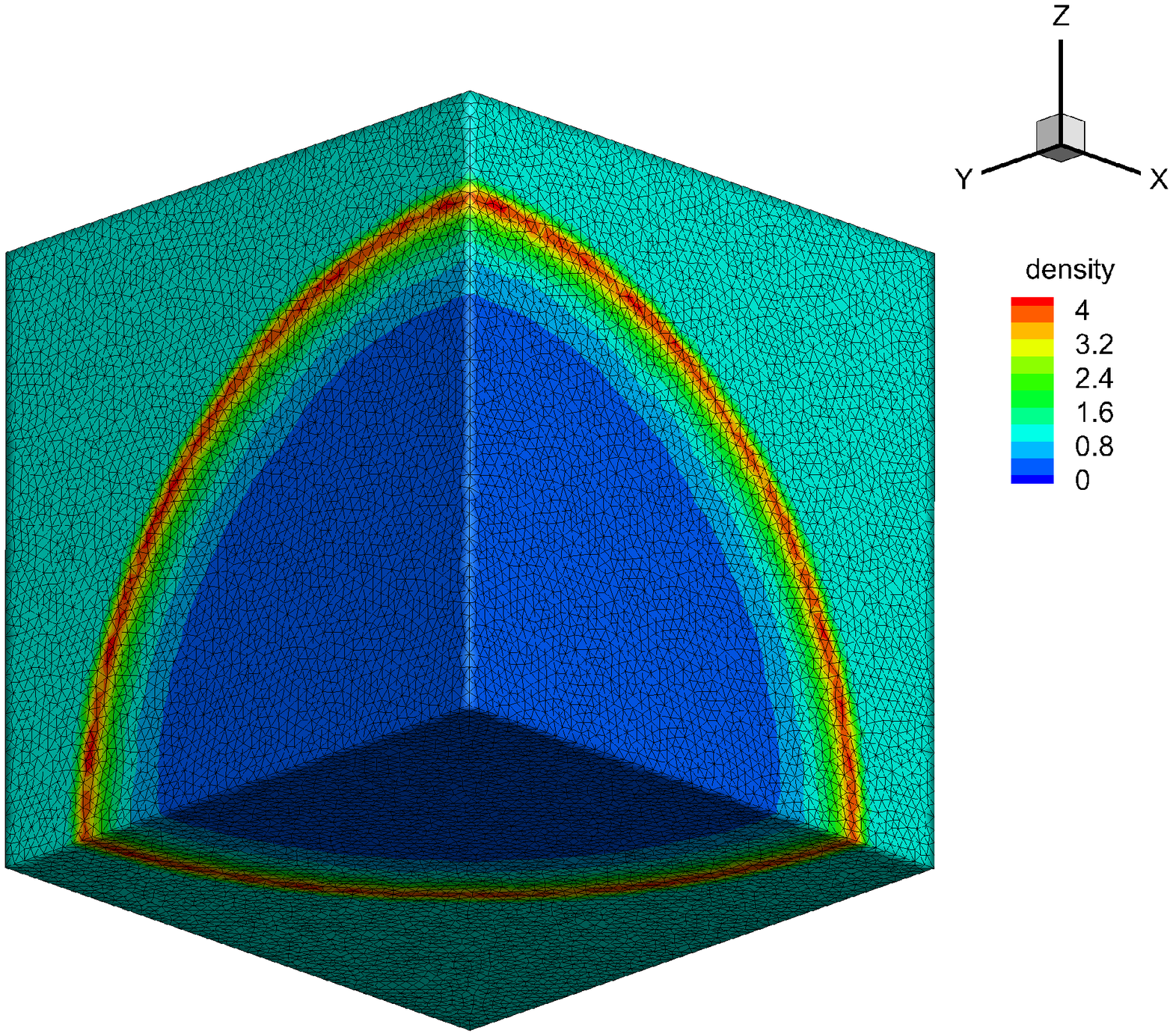}
\includegraphics[width=0.425\textwidth]{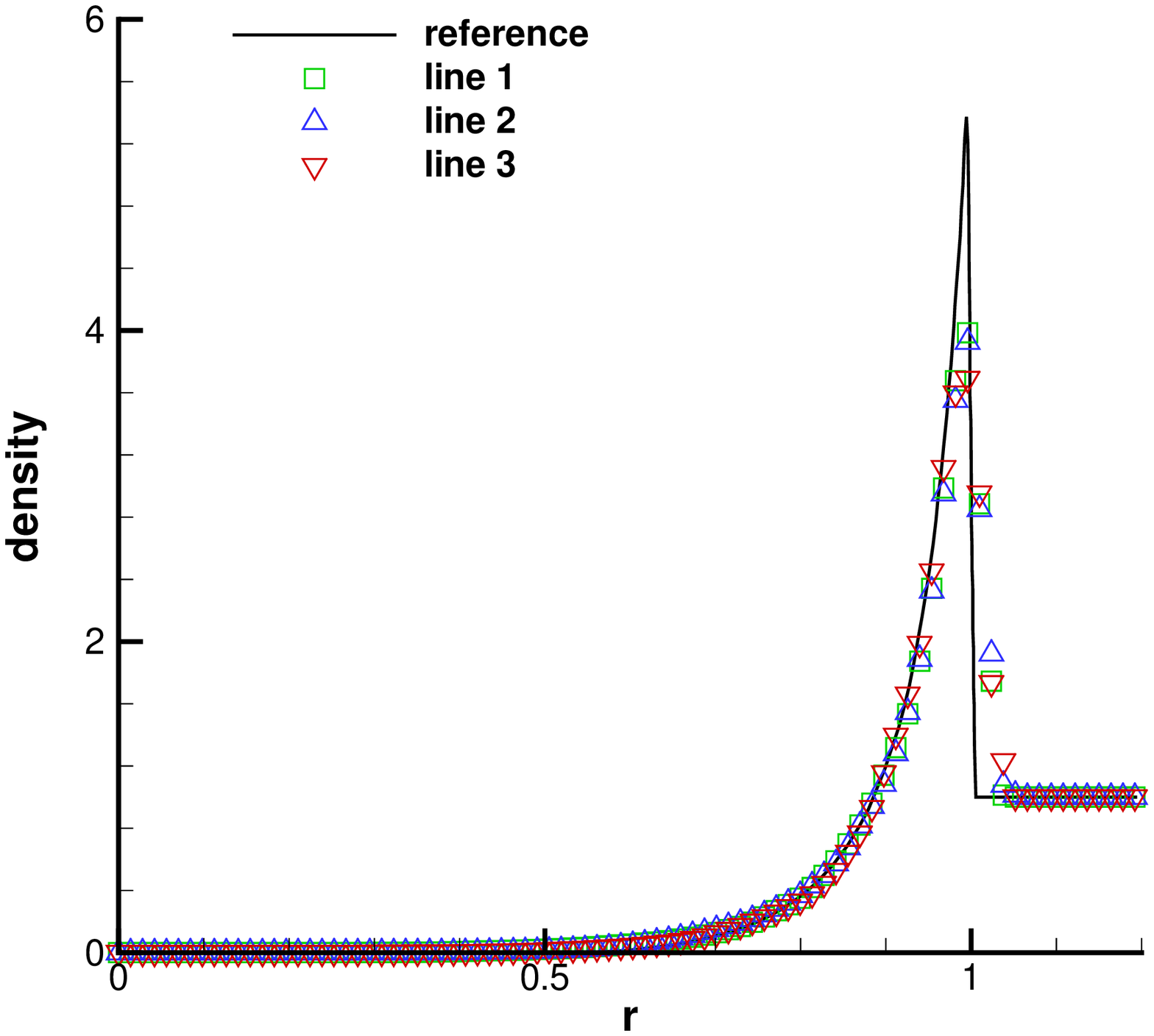}
\caption{\label{1-3d-explosion} Sedov problem: 3-D density distribution (left) and density distribution along lines (right) of Sedov problem at $t=1$. The average side length of cells of the tetrahedral mesh is $1/40$.}
\end{figure}

\begin{figure}[!htb]
\centering
\includegraphics[width=0.425\textwidth]{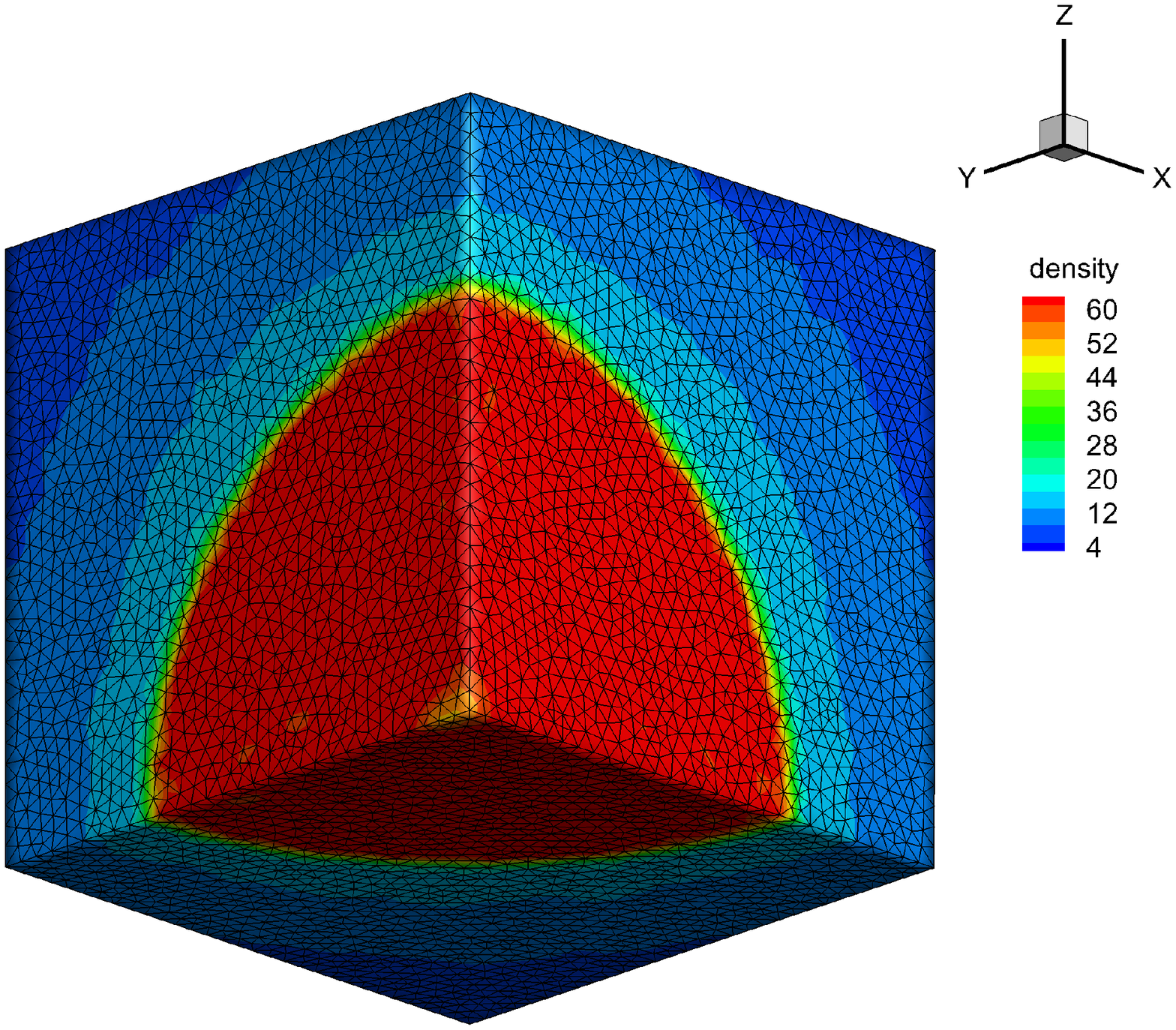}
\includegraphics[width=0.425\textwidth]{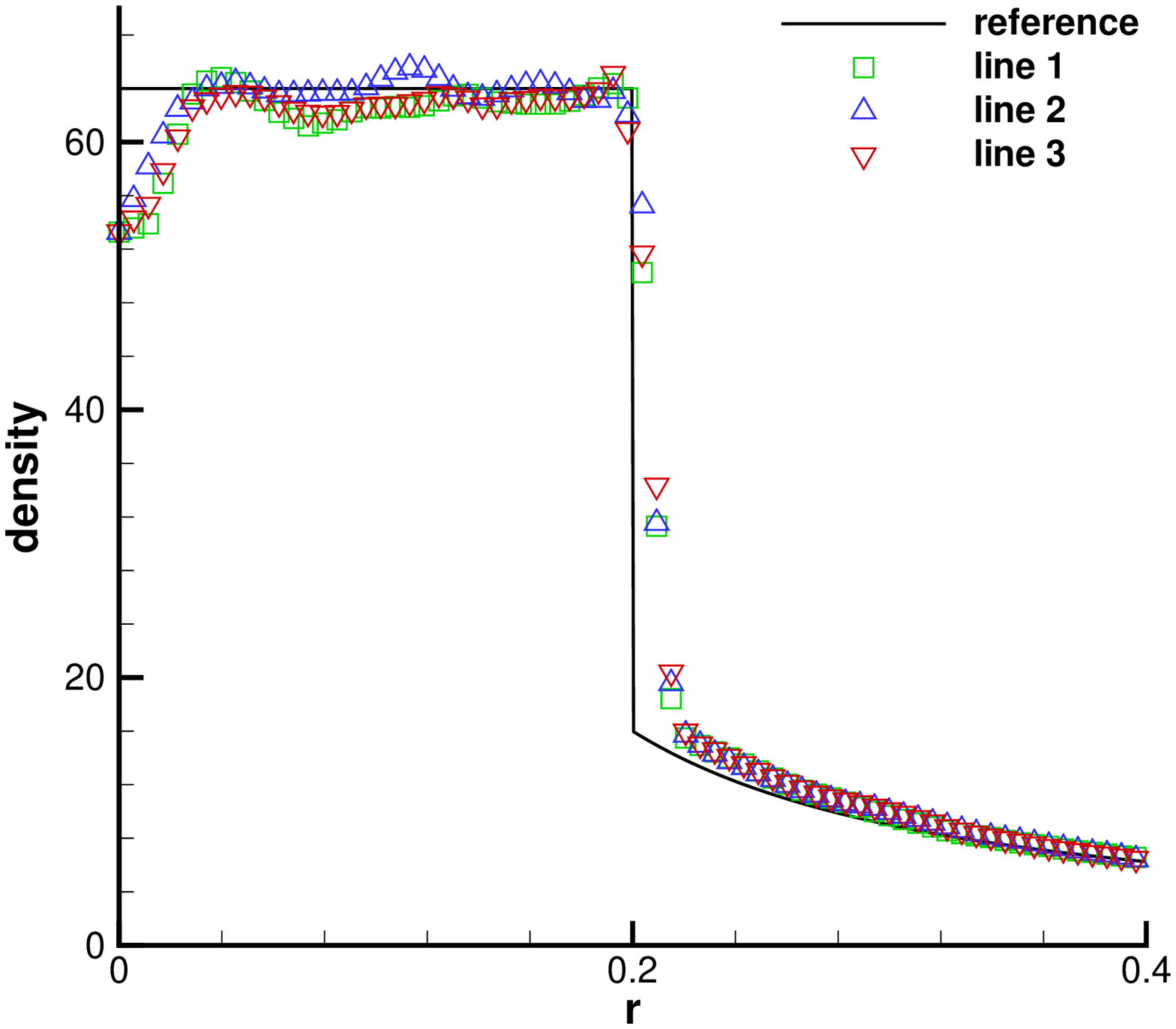}
\caption{\label{1-3d-implosion} Noh problem: 3-D density distribution (left) and density distribution along lines (right) of Noh problem at $t=0.6$. The average side length of cells of the tetrahedral mesh is $1/100$.}
\end{figure}

\subsection{3-D explosion and implosion problems}
The 3-D Sedov problem is an explosion case to model blast wave from energy deposited at a singular point \cite{sedov}.
The computational domain is $[0,1.2]^3$. The constant initial conditions with $\rho=1$, $p=1\times10^{-4}$ and $U=V=W=0$ are imposed in the whole domain except the cells containing the origin. The pressure of the cells containing the origin is set as $p=(\gamma-1)\epsilon /V$, where $\epsilon=0.106384$ and $V$ is the total volume of those cells.
The inviscid wall condition is adopted along the boundaries $x=0$, $y=0$ and $z=0$. The free boundary condition is imposed to the other boundaries.
The mesh cell size is $1/40$ defined by the average side length of tetrahedral cells. The computational output time is $t=1.0$.

The 3-D density distribution and density distribution along lines are shown in Fig \ref{1-3d-explosion}, where the reference solution is from \cite{sedov}.
Line 1, line 2 and line 3 are determined by connecting the origin to $(0,1.2,1.2)$, $(1.2,0,1.2)$ and $(1.2,1.2,0)$.
The strong robustness of the 4th-order compact GKS is demonstrated by its use of a large CFL number CFL$=0.6$ without additional limiting technique. In addition, the high resolution of the compact GKS for strong shock waves is verified by the high post-shock density peak, and the numerical shock wave remains sharp and spans only two mesh cells. The non-compact high-order GKS gives a numerical shock wave that is wider and has a lower post-shock peak \cite{pan_2021pof}.

The 3-D Noh problem is an implosion test to model the gas compression with constant radial velocity towards a spherical center, where a moving strong shock wave is formed \cite{Noh}.
The computational domain is $[0,0.3]^3$. The initial density and pressure are $\rho=1$ and $p=1\times10^{-4}$, and the velocity is $(U,V,W)=(-x,-y,-z)/\sqrt{x^2+y^2+z^2}$. The ratio of the specific heat is $\gamma=5/3$.
The inviscid wall condition is adopted along the boundaries $x=0$, $y=0$ and $z=0$. The supersonic inflow boundary condition is imposed to the other boundaries with the same pressure and velocity as the initial conditions and the analytical solution of density,
\begin{equation*}
\rho = \begin{cases}
64,  r<t/3,\\
(1+t/r)^2,  r>t/3.
\end{cases}
\end{equation*}
The mesh cell size is $1/100$ defined by the average side length of tetrahedral cells.
The final computational output time is $t=0.6$.

The 3-D density distribution and density distribution along lines are shown in Fig \ref{1-3d-implosion}.
Line 1, line 2 and line 3 are determined by connecting the origin to $(0,0.3,0.3)$, $(0.3,0,0.3)$ and $(0.3,0.3,0)$.
Again, in this test case, a large CFL number $=0.6$ without additional limiting technique is used. The accuracy of the compact GKS for strong shock waves is verified by the precise post-shock density solution which is comparable to the result of the non-compact 5th-order GKS on structured mesh \cite{yang2022comparison}.

\begin{figure}[!htb]
\centering
\includegraphics[width=0.45\textwidth]{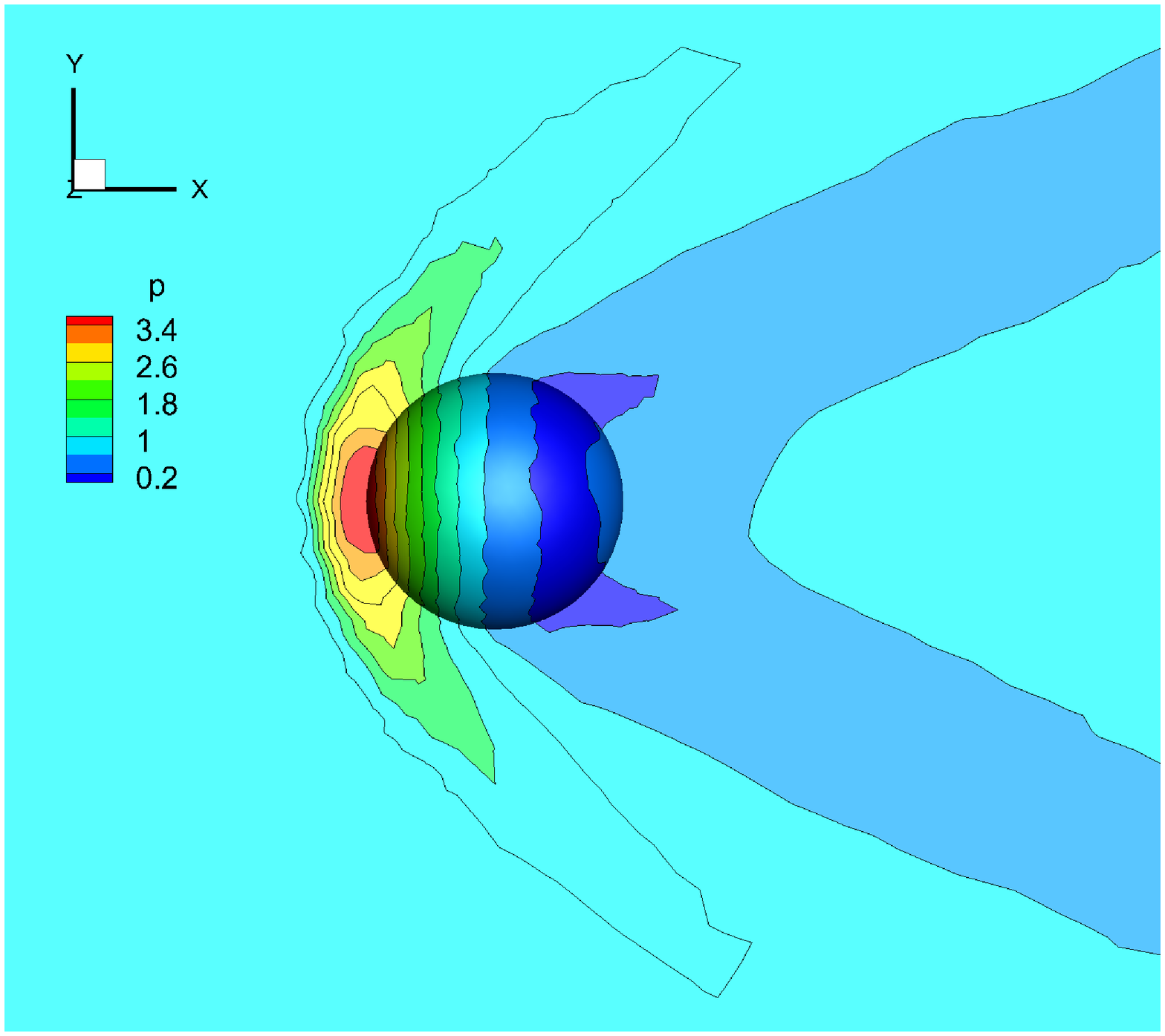}
\includegraphics[width=0.45\textwidth]{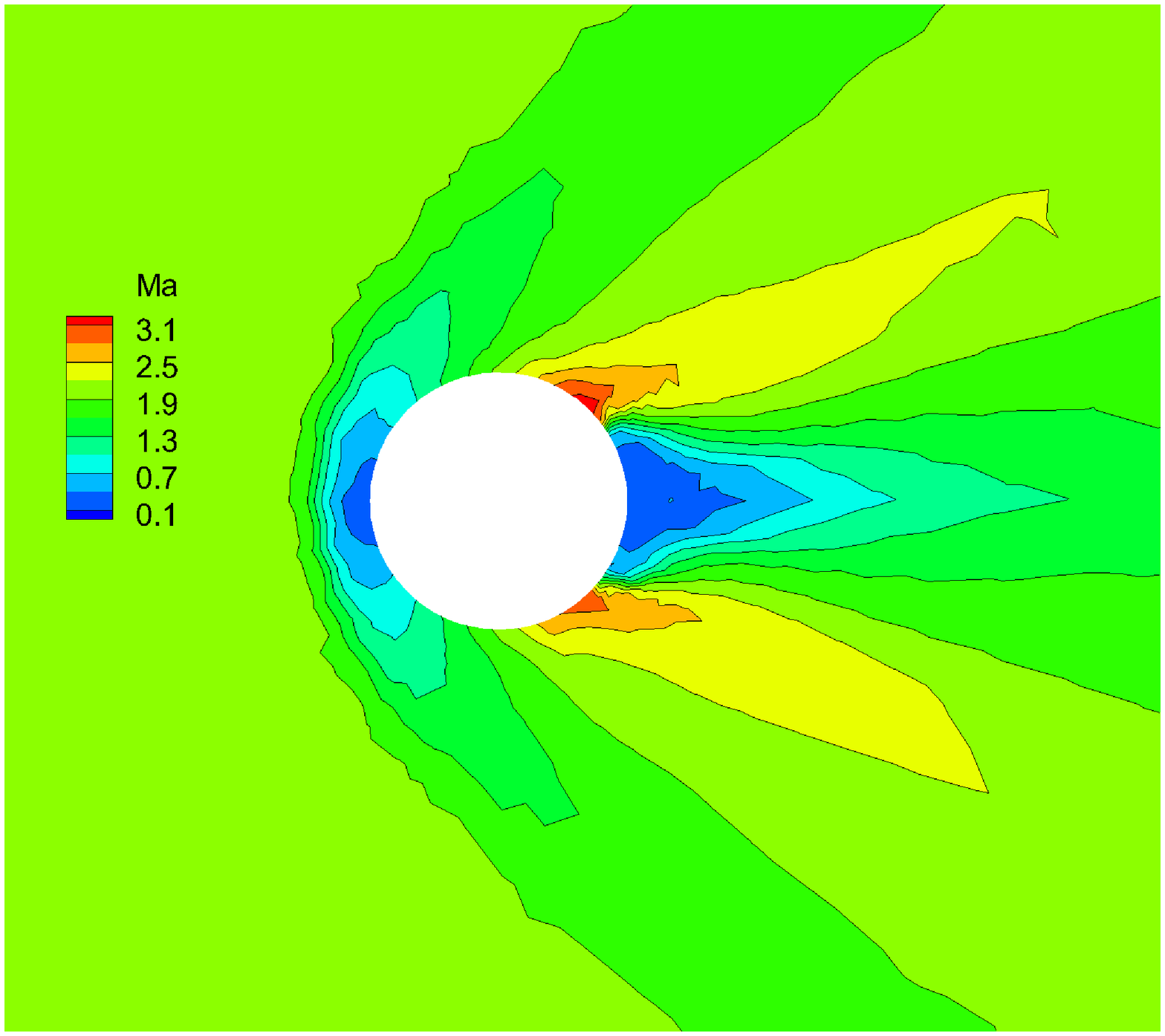}  \\
\includegraphics[width=0.45\textwidth]{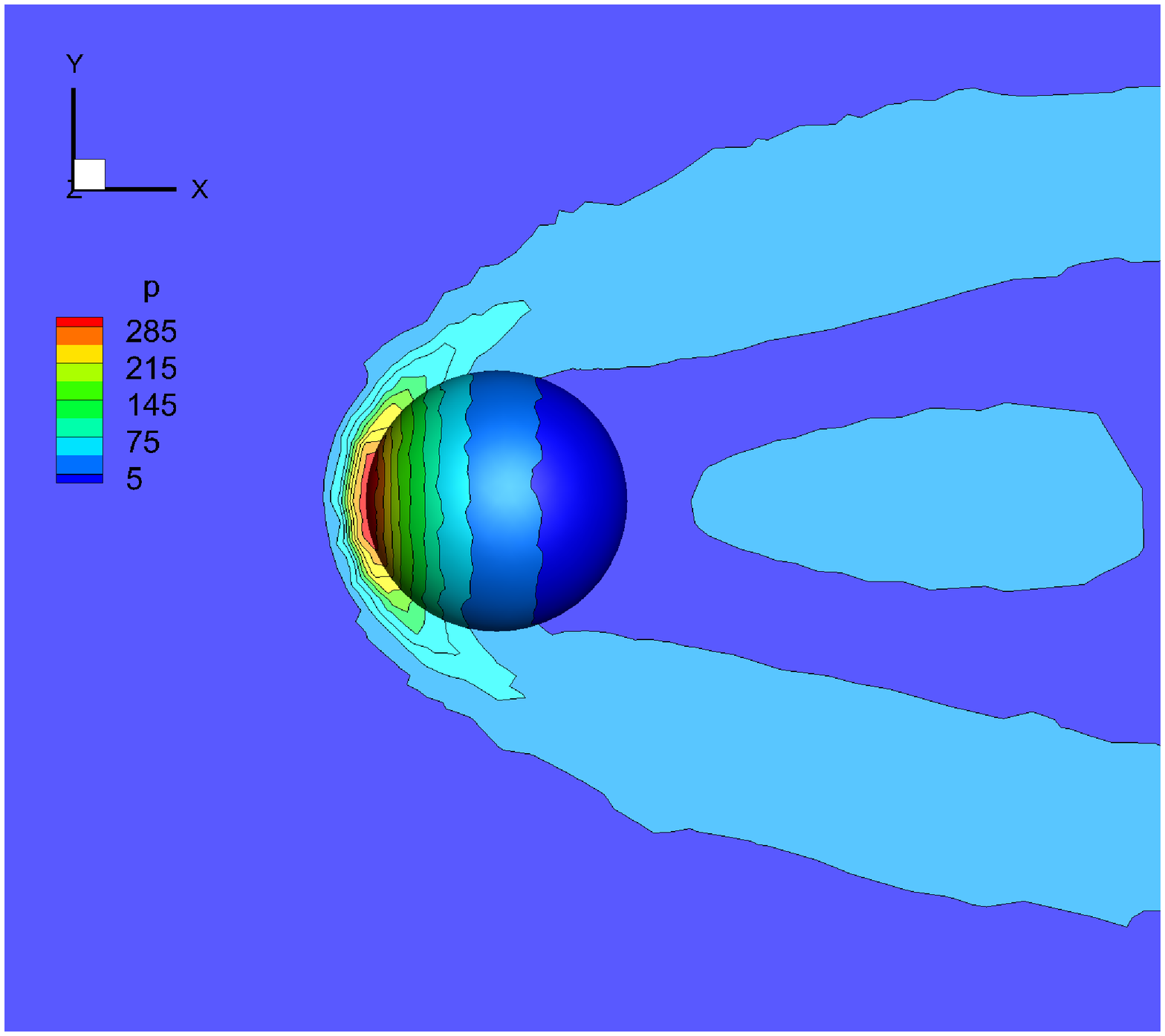}
\includegraphics[width=0.45\textwidth]{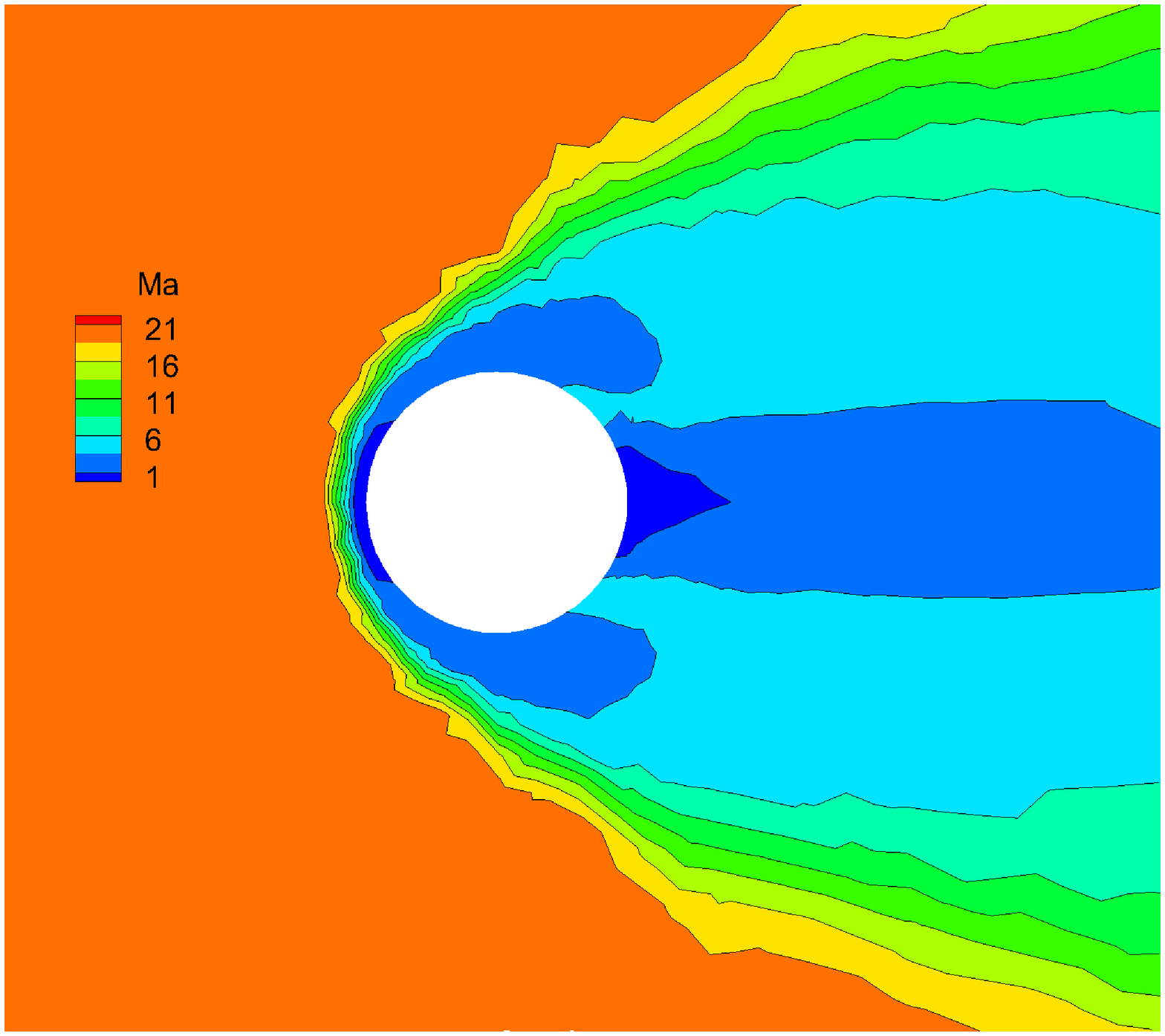}
\caption{\label{2-sphere-highmach} Supersonic and hypersonic inviscid flow around a sphere: the 3-D pressure contours (left) and 2-D Mach contours on $z=0$ plane (right) of supersonic flow with $Ma=2$ (up) and hypersonic flow with $Ma=20$ (down). The fourth-order compact GKS is used. The mesh used in the computation is the same as shown in Fig. \ref{2-vis-sphere-1}.}
\end{figure}

\subsection{Supersonic and hypersonic inviscid flow around a sphere}
The inviscid flow at high Mach numbers around a sphere is used to verify the strong robustness of the fourth-order compact GKS.
The incoming inviscid flows have Mach numbers $Ma=2$ and $Ma=20$ separately. The adiabatic reflective boundary condition is imposed on the wall of the sphere.
The same computational mesh as shown in Fig. \ref{2-vis-sphere-1} is used in the current computation.
The 3-D pressure and 2-D Mach number distributions are presented in Fig. \ref{2-sphere-highmach} when the flow reaches steady state.
In this test case, the exact same code and parameter settings are used as the nonlinear reconstruction-based GKS in the accuracy test
of Fig. \ref{Accuracy-time}.

\begin{figure}[!htb]
\centering
\includegraphics[width=0.495\textwidth]{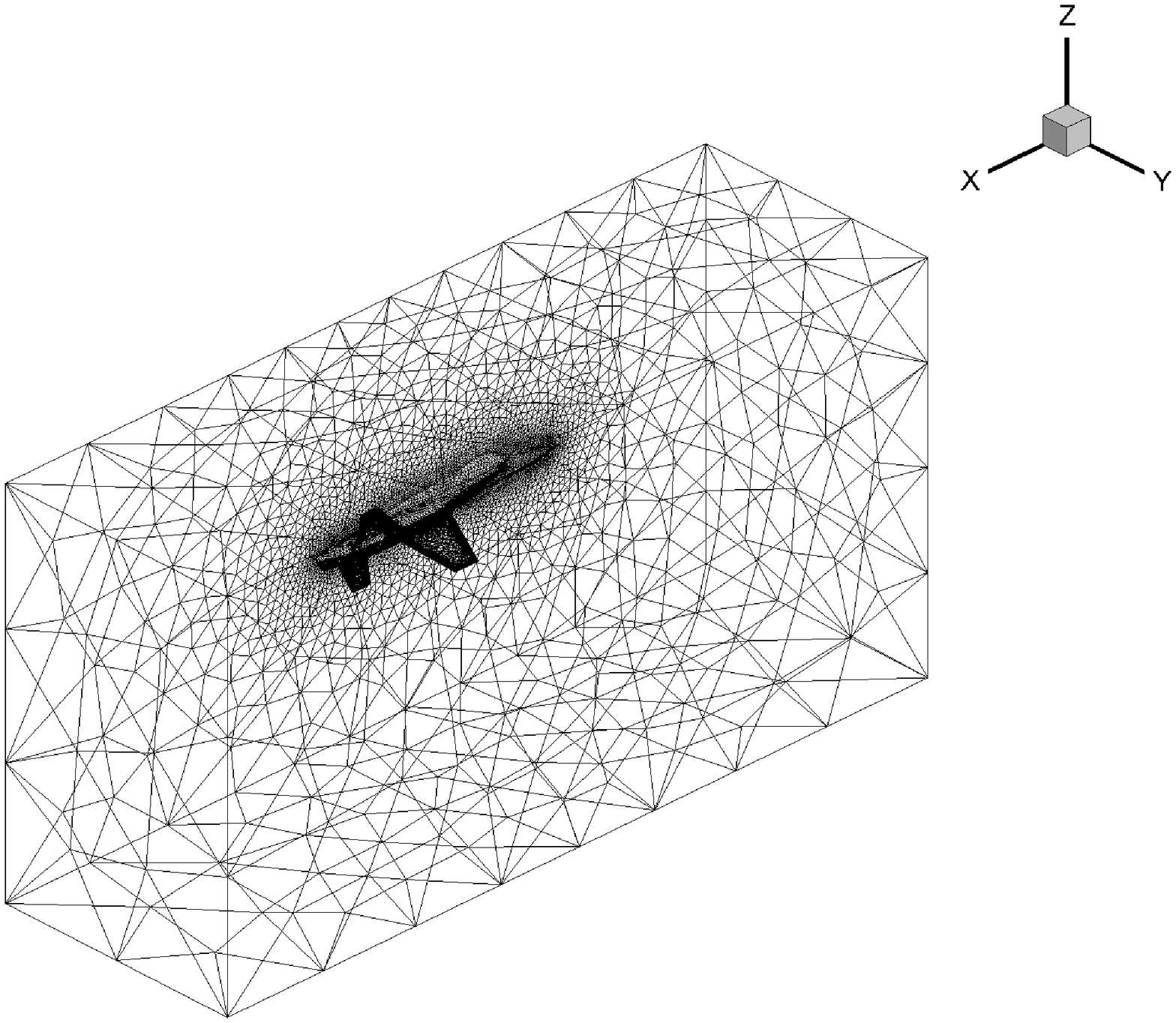}
\includegraphics[width=0.495\textwidth]{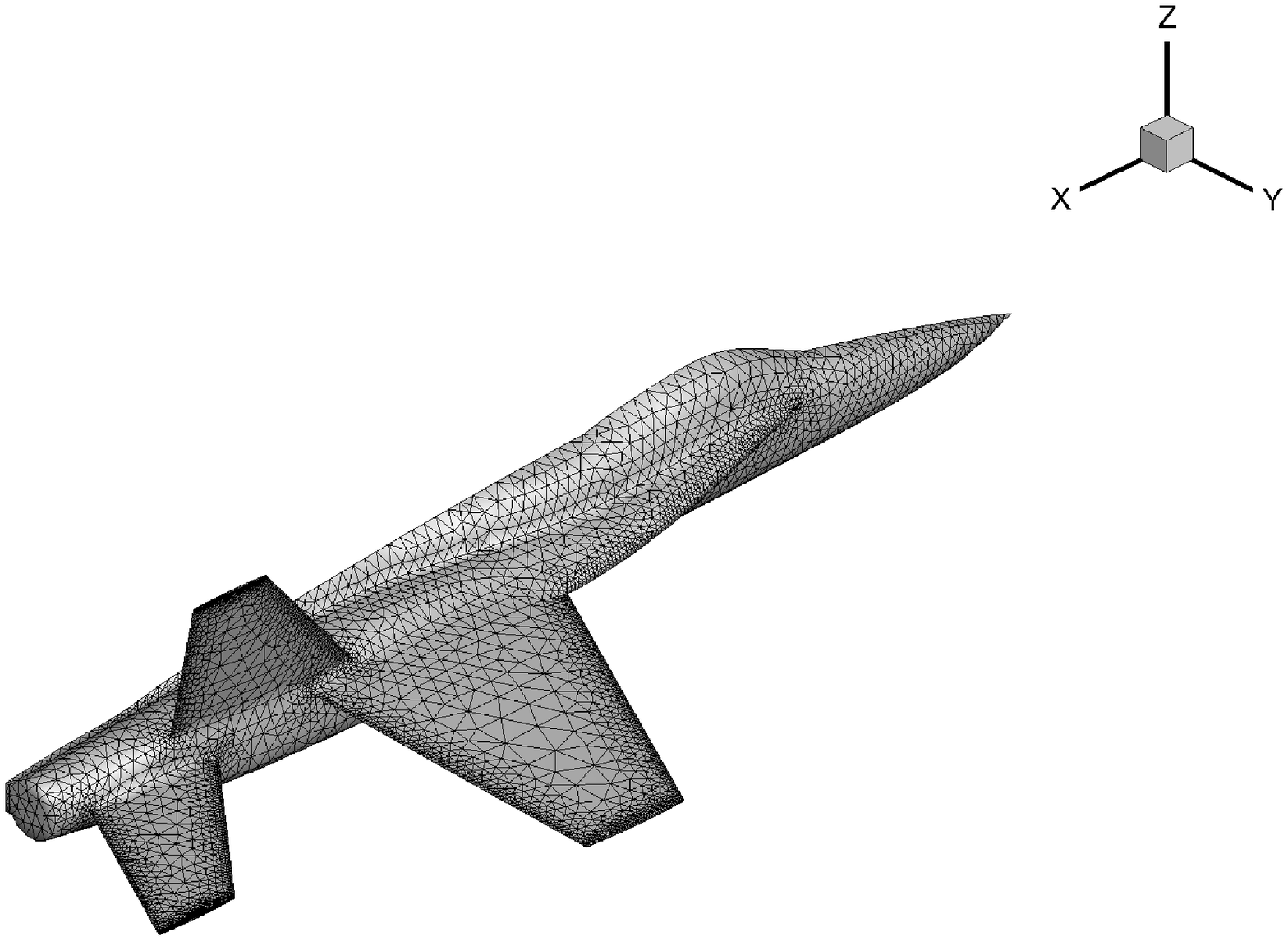}
\caption{\label{2-fighter-model-1} Supersonic flow over the YF-17 model: the mesh used in the computation. }
\end{figure}

\begin{figure}[!htb]
\centering
\includegraphics[width=0.495\textwidth]{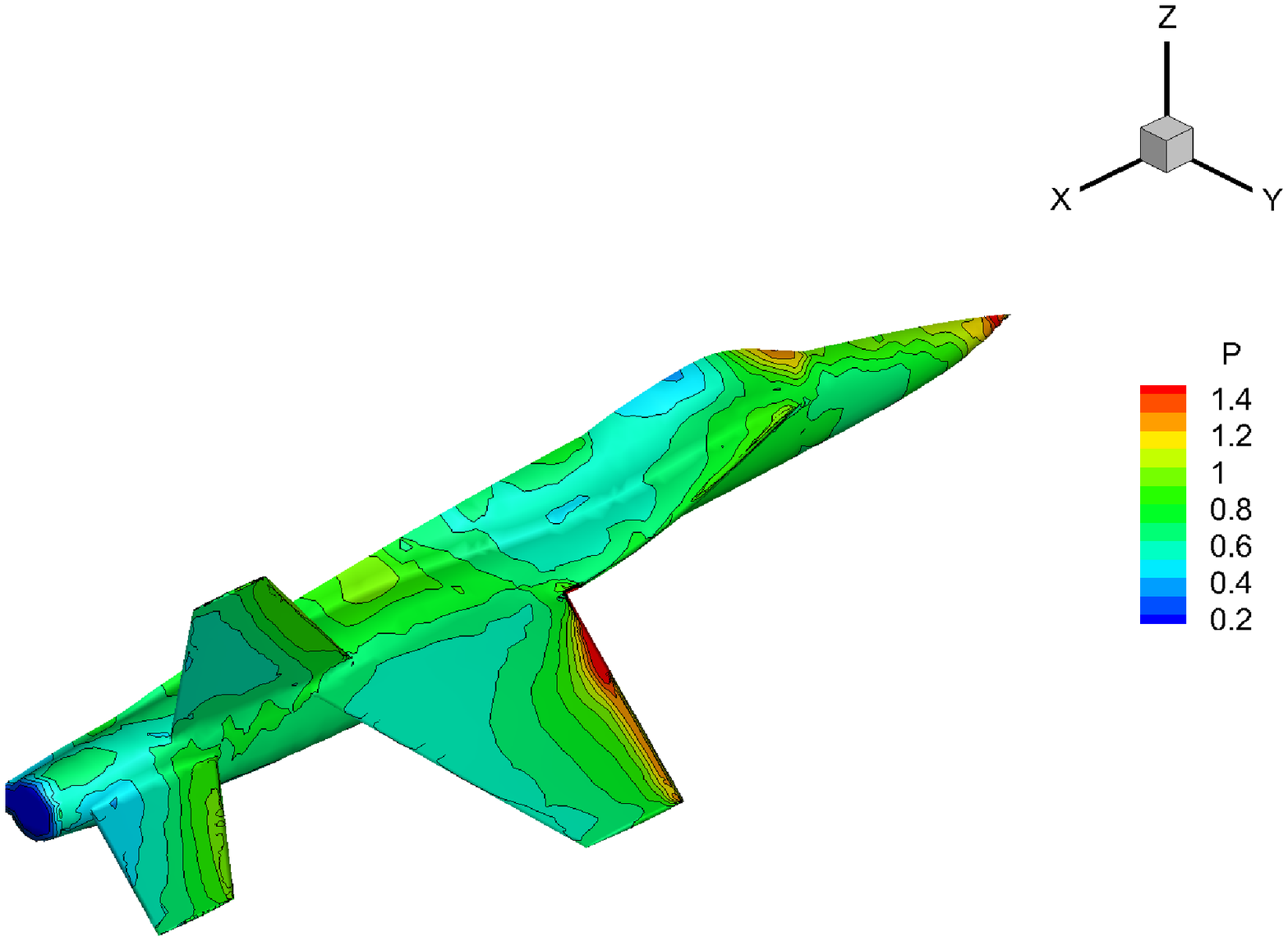}
\includegraphics[width=0.495\textwidth]{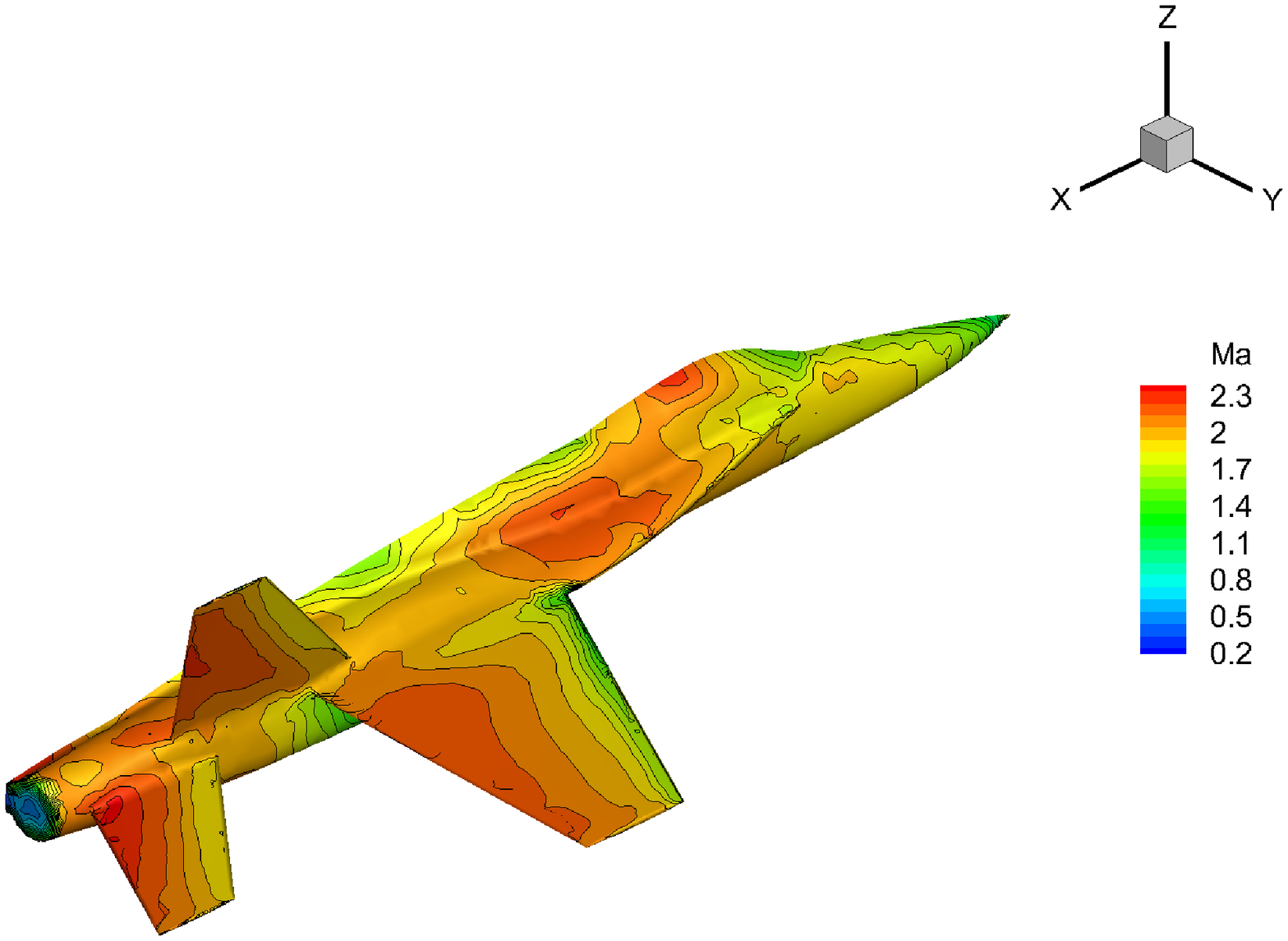}\\
\includegraphics[width=0.495\textwidth]{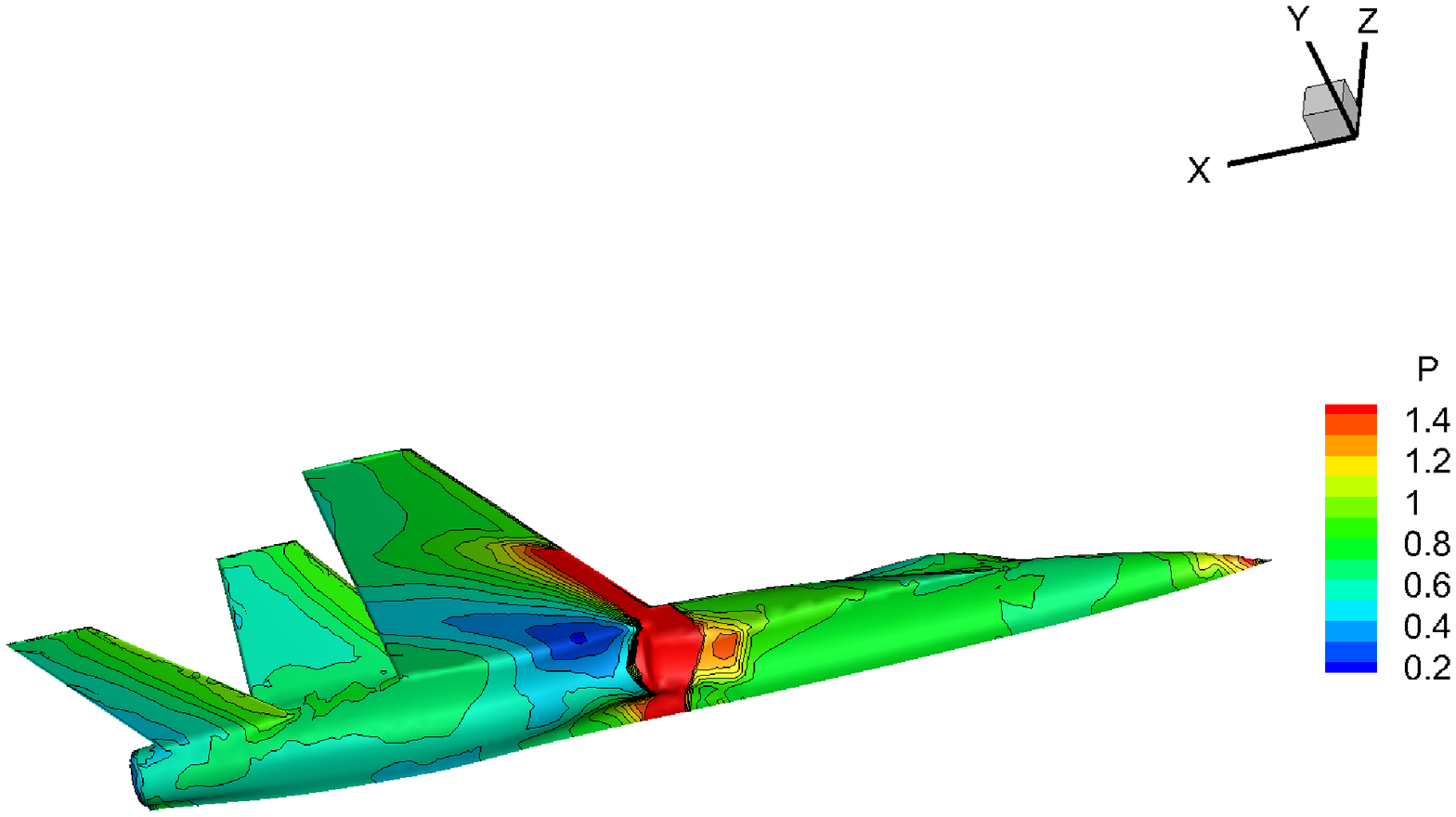}
\includegraphics[width=0.495\textwidth]{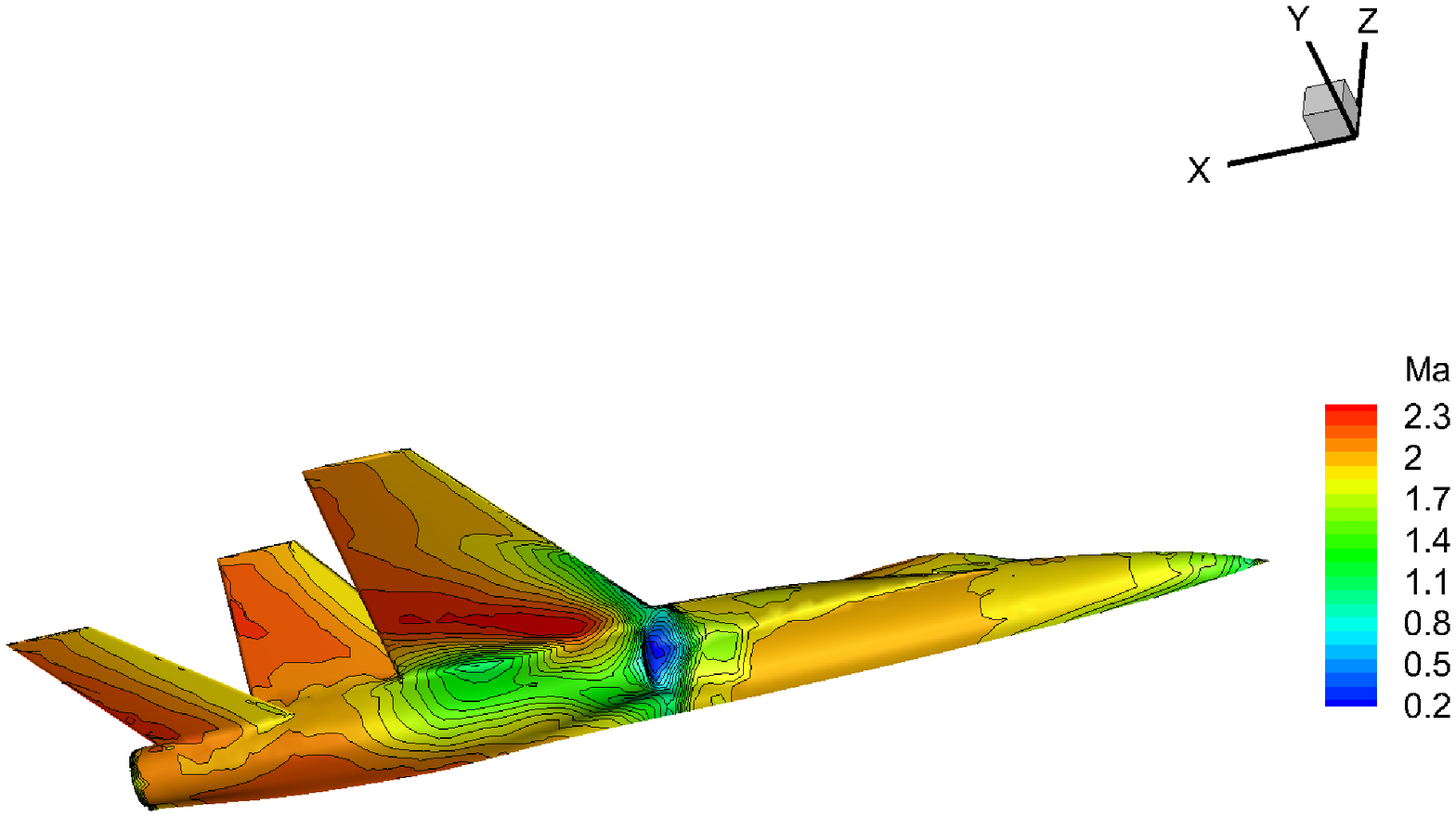}
\caption{\label{2-fighter-model} Supersonic flow over the YF-17 model: the results of steady state obtained by 4th-order compact GKS. The pressure and Mach number contours of top view (up) and down view (down).}
\end{figure}

\subsection{Supersonic flow over the YF-17 fighter model}
The supersonic flow over the YF-17 model is simulated by the fourth-order GKS under complex geometry.
The mesh of the YF-17 model is provided at {\sl https://cgns.github.io/CGNSFiles.html}, which is shown in Fig. \ref{2-fighter-model-1}.
The total number of tetrahedral cells is about $1.7\times 10^5$.
The incoming Mach number is set as $Ma_{\infty}=2.0$, and the angle of attack is $AoA=0$. The inviscid wall boundary condition is adopted on the model surface and on the symmetry plane.

Fig. \ref{2-fighter-model} presents the pressure and Mach number distributions on the model at a steady state.
A low pressure area occurs in the downwind area at the extreme end of the  model fuselage. Shock waves are generated at the nose of the model fuselage and the front of the wing.

\section{Conclusion}
In this paper, high-order compact GKS from second- up to fourth-order have been constructed on three-dimensional tetrahedral mesh for compressible flow simulations.
The compact scheme works very well from the subsonic smooth flow to the hypersonic compressible flow simulations.
The scheme shows the accuracy/effieicency in the smooth viscous flow computation and excellent robustness for the complex flow with discontinuities. More importantly, a large CFL number, such as CFL=0.6, can be used for the fourth-order compact GKS, even at a Mach number $20$ flow computation and Noh problem on tetrahedral mesh.
The success of the compact scheme is mainly coming from following fact in the algorithm development.
A high-order time-accurate gas evolution model at a cell interface is adopted and it provides the evolution solution of flow variables
and fluxes at cell interfaces, which can be used to update cell-averaged flow variables and their gradients.
More importantly, in the high-order evolution model a nonlinear limiter on the high-order time derivative of flux function
through WENO formulation is introduced in its temporal evolution, which makes the status of ``nonlinear limiters" on the equivalent footing in the spatial reconstruction and temporal evolution for capturing
the propagation of discontinuous solution in space and time.
As a result, even for the fourth-order compact scheme, a large CFL number can be used, such as CFL$=0.6$ in the current scheme in comparison with the $0.14$ in the DG method.
At the same time, based on the cell-averaged flow variables and their gradients improved WENO weighting functions in the compact reconstruction have been  developed.
The extension of the current compact GKS to three-dimensional mixed unstructured mesh is straightforward.
It is expected that the high-order compact GKS will play an important role in engineering applications.

\section{Acknowledgments}

The authors would like to thank Dr. Yajun Zhu for helpful discussion on the three-dimensional rotational coordinate transformation and Mr. Junzhe Cao for providing mesh used in computation.
The current research is supported by National
Science Foundation of China (No.12172316), Hong Kong research grant council 16208021 and 16301222,
and CORE as a joint research centre for ocean research between QNLM and HKUST through the project QNLM20SC01-A and QNLM20SC01-E.

\section*{Appendix}
The rotational coordinate transformation for the variables and their derivatives are required. For example, the reconstructed flow variables are given as $Q(\mathbf{x})$ and $\nabla Q(\mathbf{x})$ in the coordinate system $x-y-z$, then the variables in the local coordinate system $\widetilde{x}-\widetilde{y}-\widetilde{z}$ are needed for the calculations of flow variables and numerical fluxes at cell interface, where the positive direction of $\widetilde{x}$ is the same as the unit outer normal vector $\mathbf{n}$ of the cell interface. The two coordinate systems can be linked by a rotation transformation matrix.
The rotational transformation from the coordinate system $x-y-z$ to $\widetilde{x}-\widetilde{y}-\widetilde{z}$ is defined as
\begin{equation}\label{coor-trans}
(\widetilde{x},\widetilde{y},\widetilde{z})^\mathrm{T}=\mathbf{M} (x,y,z)^\mathrm{T},
\end{equation}
where $\mathbf{M}$ is the rotational matrix. $\mathbf{M}$ can be given as $\mathbf{M}=(\mathbf{M}_1,\mathbf{M}_2,\mathbf{M}_3)$, where $\mathbf{M}_k~(k=1,2,3)$ is the row vector.
A simple algorithm to determine $\mathbf{M}_k~(k=1,2,3)$ is as follows.
\begin{enumerate}
    \item Given the unit outer normal vector $\mathbf{n}=(n_1,n_2,n_3)$ of the cell interface, and find $n_k$ which has the smallest absolute value.
    \item Define unit vector $\mathbf{n}^a$ whose components are $0$ except for $n^a_k=1$.
    \item Calculate $\mathbf{n}^b=(\mathbf{n}\cdot \mathbf{n}^a)\mathbf{n}$, and $\mathbf{t}^{a}=\frac{\mathbf{n}^a-\mathbf{n}^b}{|\mathbf{n}^a-\mathbf{n}^b|}$.
    \item calculate $\mathbf{t}^{b}=\mathbf{n} \times \mathbf{t}^{a}$.
    \item $\mathbf{M}_1=\mathbf{n}$, $\mathbf{M}_2=\mathbf{t}^{a}$, $\mathbf{M}_3=\mathbf{t}^{b}$.
\end{enumerate}

The conservative variables $\widetilde{\mathbf{W}}(\widetilde{\mathbf{x}})$ in the coordinate system $\widetilde{x}-\widetilde{y}-\widetilde{z}$ can be obtained from $\mathbf{W}(\mathbf{x})$ in the coordinate system $x-y-z$ by
\begin{equation}\label{G-to-L}
\widetilde{\mathbf{W}}(\widetilde{\mathbf{x}})=\mathbf{T} \mathbf{W}(\mathbf{x}),
\end{equation}
where matrix $\mathbf{T}$ is the rotational matrix. $\mathbf{T}$ is an orthogonal identity matrix and is given by
\begin{equation}\label{coor-trans-T}
\textbf{J}={
	\left( \begin{array}{ccc}
	1 & \mathbf{0}              &0 \\
	\mathbf{0} & \mathbf{M}     &\mathbf{0} \\
    0 & \mathbf{0}              &1
	\end{array}
	\right ).}
\end{equation}
$\nabla \widetilde{\mathbf{W}}(\widetilde{\mathbf{x}})$ in the coordinate system $\widetilde{x}-\widetilde{y}-\widetilde{z}$ can be obtained from $\nabla \mathbf{W}(\mathbf{x})$ in the coordinate system $x-y-z$ by
\begin{equation}\label{G-to-L-der}
\begin{split}
&(\nabla_{\widetilde{k}}\widetilde{\mathbf{W}})^T=\mathbf{M} (\nabla_{k}\widetilde{\mathbf{W}})^T,\\
&\partial_{k}\widetilde{\mathbf{W}}=\mathbf{T}\partial_{k}\mathbf{W}.
\end{split}
\end{equation}
where $k=x,~y,~z$.
The inverse transformation of Eq.(\ref{G-to-L}) and Eq.(\ref{G-to-L-der}) can be easily obtained since $\mathbf{T}$ is an orthogonal identity matrix.

\bibliographystyle{ieeetr}
\bibliography{AIAbib}

\end{document}